\documentclass[usenatbib]{mnras}
\usepackage[T1]{fontenc}
\usepackage{aecompl}
\usepackage{amsmath,amsfonts,amssymb}
\usepackage[pdftex]{graphicx}
\usepackage{aas_macros}
\usepackage[usenames]{xcolor}
\usepackage{color}
\usepackage{textcomp,gensymb}
\usepackage{times}
\usepackage{subfig}

\title[Flybys in dusty protoplanetary discs]{Flybys in protoplanetary discs: I. Gas and dust dynamics}

\author[N. Cuello et al.]{\parbox{\textwidth}{
Nicol\'as Cuello$^{1,7}$\thanks{corresponding author: ncuello@astro.puc.cl},
Giovanni Dipierro$^{2}$,
Daniel Mentiplay$^{3}$,
Daniel~J. Price$^{3}$,
Christophe Pinte$^{3}$,
Jorge Cuadra$^{1,6,7}$,
Guillaume Laibe$^{4}$,
Fran\c cois M\'enard$^{5}$,
Pedro P. Poblete$^{1,6,7}$
and Mat\'ias Montesinos$^{7,8, 9}$}\vspace{0.2cm}\\
$^{1}$Instituto de Astrof\'isica, Pontificia Universidad Cat\'olica de Chile, Santiago, Chile,\\
$^{2}$Department of Physics and Astronomy, University of Leicester, Leicester, LE1 7RH, United Kingdom,\\
$^{3}$Monash Centre for Astrophysics (MoCA) and School of Physics and Astronomy, Monash University, Clayton, Vic 3800, Australia,\\
$^{4}$Univ Lyon, Univ Lyon1, Ens de Lyon, CNRS, Centre de Recherche Astrophysique de Lyon UMR5574, F-69230, Saint-Genis-Laval, France,\\
$^{5}$Univ. Grenoble Alpes, CNRS, IPAG, F-38000 Grenoble, France,\\
$^{6}$Millennium Nucleus ``Protoplanetary Disks", Santiago, Chile, \\
$^{7}$N\'ucleo Milenio de Formaci\'on Planetaria (NPF), Chile, \\
$^{8}$Instituto de F\'isica y Astronom\'ia, Universidad de Valpara\'iso, Chile, \\
$^{9}$Chinese Academy of Sciences South America Center for Astronomy, National Astronomical Observatories, CAS, 
Beijing 100012, China.
}

\pagerange{\pageref{firstpage}--\pageref{lastpage}} \pubyear{2018}
\date{}
\begin{document}
\label{firstpage}
\date{Accepted ... Received ...}

\maketitle

\begin{abstract}
We present 3D smoothed particle hydrodynamics simulations of protoplanetary discs undergoing a flyby by a stellar perturber on a parabolic orbit lying in a plane inclined relative to the disc mid-plane. We model the disc as a mixture of gas and dust, with grains ranging from 1 $\mathrm{\mu}$m to 10 cm in size. Exploring different orbital inclinations, periastron distances and mass ratios, we investigate the disc dynamical response during and after the flyby. We find that flybys induce evolving spiral structure in both gas and dust which can persist for thousands of years after periastron. Gas and dust structures induced by the flyby differ because of drag-induced effects on the dust grains. Variations in the accretion rate by up to an order of magnitude occur over a time-scale of order 10 years or less, inducing FU~Orionis-like outbursts. The remnant discs are truncated and warped. The dust disc is left more compact than the gas disc, both because of disc truncation and accelerated radial drift of grains induced by the flyby.
\end{abstract}

\section{Introduction}
\label{sec:intro}

Star formation is chaotic and violent, with interaction between young stars common in molecular clouds \citep{Bate2018}. Planets form in discs of gas and dust orbiting these stars. The presence of other stars in the field can dramatically affect the disc morphology \citep{Pfalzner2003,2008ASPC..398..341C, Vinke+2015, Dai+2015, XG2016, Winter+2018b}. In this paper, we consider how dynamical interactions with companions on unbound orbits affect the structure and evolution of protoplanetary discs, and the implications for planet formation.

Stellar encounters are expected to occur mainly during the first Myr of evolution of star clusters. \cite{Pfalzner2013} estimated the probability for a star to undergo such an encounter --- defined by having a periastron distance in the range $[100,1000]$ au --- by considering the mass and eccentricity distributions of encounters for solar-type stars --- mass in a range $[0.8,1.2]\,M_{\odot}$ --- in the so-called ``leaky'' clusters (OB associations). Based on the stellar cluster properties, she found that roughly 30\% of the solar-type stars experience an encounter during the early evolutionary stages. The recent work by \cite{Winter+2018b} who considered different models of stellar clusters is consistent with this estimate (cf. their figure 7). Given that the probability of an encounter rapidly decreases with time (as the stellar density), if an encounter occurs, then it is statistically most likely to occur during the protostellar stage. In addition, encounters with low or solar-mass companions were found to be more probable than encounters with companions of several $M_\odot$ up to $100$ $M_\odot$. Since the perturber is unbound from the host star, its eccentricity may be equal  to or larger than unity, with values typically between $1$ and $10$, reaching $100$ for the most extreme cases (figures~5, 6 and 7 in \citealt{Pfalzner2013}).

In a seminal paper, \cite{Clarke&Pringle1993} considered coplanar parabolic encounters between equal-mass stars with periastron separations of the order of the initial disc size. Prograde encounters were found to be the most destructive, tidally truncating the disc and unbinding material that is either captured by the perturber or escapes \citep{Breslau+2017}. Retrograde encounters, by contrast, were found to leave the disc intact. Interestingly, for polar (orthogonal) orbits there is almost no captured material by the intruder but the disc is significantly warped \citep{nixon10a}.

Subsequent authors examined three main effects caused by stellar flybys: i) tidal truncation of the disc \citep{Breslau+2014,Bhandare+2016,Breslau+2017}; ii) generation of a highly inclined disc \citep{Terquem&Bertout1993, Terquem&Bertout1996, XG2016}; and iii) fragmentation of the disc due to self-gravity \citep{Thies2010}. 
 
\cite{Breslau+2014,Breslau+2017} and \cite{Bhandare+2016} used $N$-body calculations to study tidal truncation. \citet{Breslau+2017} computed the final disc size as a function of the periastron and the mass ratio, assuming prograde coplanar parabolic encounters. They also estimated the portion of unbound, captured and remaining material around the primary star after the flyby. They found two main regimes --- corresponding to low- and high-mass perturbers --- with a wide distribution in the final orbital properties of particles. The equal-mass case was found to be a combination of both regimes (see figures~2, 3 and 5 in \citealt{Breslau+2017}). They also found that disc sizes may have been over-estimated in the past by the mis-application of equal-mass results from \citet{Clarke&Pringle1993} to non-equal mass encounters. \cite{Bhandare+2016} proposed a more accurate formula to estimate the final size of the disc after a parabolic inclined stellar flybys (cf. Sect.~\ref{sec:truncation}). They also showed that the strength of the tidal interaction for unbound orbits is stronger when the inclination and the eccentricity are small. However, effects related to disc viscosity, thermal pressure, self-gravity and vertical geometry were not accounted for.
 
\citet{Terquem&Bertout1993, Terquem&Bertout1996} and \citet{XG2016} showed that stellar flybys on inclined orbits could generate highly inclined discs. Again, retrograde orbits were found to be less destructive than prograde orbits. Inclined retrograde orbits were found to generate disc inclinations of up to $60\degree$ with respect to the binary orbit. Perturbers on inclined prograde orbits were not found to produce such high disc inclinations, mainly because more material is captured by the perturber. \citet{XG2016} demonstrated that the final disc orientation is determined by the precession of the disc angular momentum about the perturber's orbital momentum vector. Again, smaller periastron distances and higher mass companions were found to induce the strongest effects.

\cite{Thies2010} showed that flybys could trigger gravitational instability. In massive gaseous discs of roughly 0.5 $M_\odot$, they showed that a flyby could lead to brown dwarf and giant-planet formation in the outer regions of the disc. This provides a way to form massive companions/planets at large distances from the host star. However, in gravitationally unstable discs, fragmentation can be inhibited by the tidal heating released during the interaction as shown by \cite{Lodato+2007}. This effect tends to stabilize the disc and strongly depends on the disc cooling rate, the flyby geometry and the perturber's properties. 

Our study is motivated by the increasing number of discs that are observed to be warped, misaligned or characterized by an asymmetric structure in emission \citep[e.g.][]{Rosenfeld+2012,grady13a,benisty15a,stolker16a,Langlois+2018,casassus18a,boehler18a}. The case of RW Aur is of particular interest since striking discrepancies have been observed between the gas and the dust distributions (e.g. \citealt{Dai+2015}). According to \cite{Rodriguez+2018}, a recent flyby which left peculiar fingerprints both in the dust and in the gas is likely. More generally, with the spectacular advances in resolution and sensitivity of ALMA and current adaptive optics instruments (e.g. SPHERE), we foresee many more discoveries in the near future. These puzzling observations challenge our understanding of disc evolution. Since observations of scattered light by dust grains and the thermal dust emission remain the key diagnostic tools to detect discs and to characterize their structure, our main aim is to predict the dynamical effect of flybys on the dust disc. No modelling currently exists for the effect of flybys in dusty discs, despite emission from $100\,\mu$m to mm-sized grains being the main tracer visible to ALMA and dust being the seed material for planet formation \citep{testi14a}.

Our approach is to study the dynamics of dust and gas during flybys using three dimensional hydrodynamics simulations of viscous dust/gas protoplanetary discs. We restrict our focus to the dynamical signatures of the flyby, both during and after the encounter, leaving detailed radiative transfer modelling of the observational signatures for a subsequent study. Our simulations cover a modest, but meaningful, portion of the allowed space of parameters: orbital inclination, periastron, mass ratio and grain size. We describe the hydrodynamical code and disc setup in Section~\ref{sec:sims}. Section~\ref{sec:results} presents our main results. Section~\ref{sec:discussion} discusses the implications for planet formation. We summarize our results in Section~\ref{sec:conclusions}.

\section{Methods}
\label{sec:sims}

We perform 3D dust/gas hydrodynamics simulations using {\sc phantom}, a smoothed particle hydrodynamics (SPH) code  \citep{Price+2017}. SPH (for reviews, see e.g. \citealt{Monaghan2005,Price2012}) is well suited to simulating flybys because there is no preferred geometry and angular momentum is conserved to the accuracy of the timestepping scheme irrespective of the plane of the orbit. {\sc phantom} has already been used widely for studies of dust and gas in discs \citep[e.g.][]{Dipierro+2015,Dipierro+2016,ragusa17a}. 

\begin{figure}
\begin{center}
\includegraphics[width=0.5\textwidth]{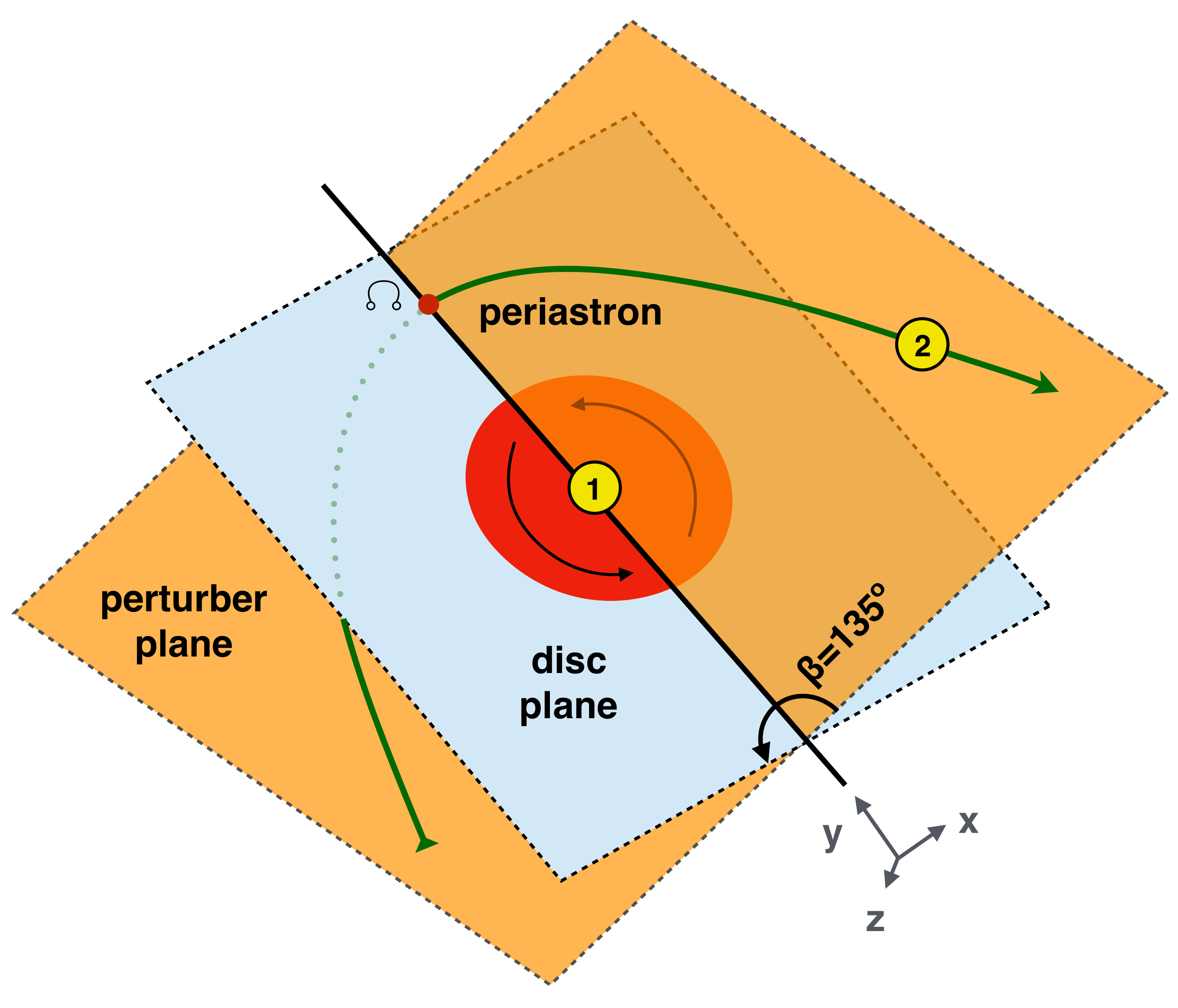}
\caption{Sketch of the inclined retrograde orbit $\beta=135\degree$. The disc is initially in the $xy$-plane (light blue). The perturber's orbit lies in an inclined plane (orange) and crosses the disc plane at periastron. From this perspective, the perturber (2) arrives from below and leaves above the disc plane.}
\label{fig:sketch}
\end{center}
\end{figure}

\subsection{Initial conditions}
\label{sec:setup}

We show our results in a reference frame centred on the primary, i.e. the star surrounded by a circumprimary disc. The reference frame in the code itself is arbitrary --- motion of the centre of mass is computed self-consistently in {\sc phantom}, avoiding the need for indirect acceleration terms in the potential as in \cite{XG2016}.

\subsubsection{Circumprimary gas disc}

We assume a circumprimary disc orbiting around a star with mass $1\,M_\odot$, with the midplane initially aligned with the $xy$-plane of our simulations. We set the disc inner and outer radii to $R_{\rm in}=10$ au and $R_{\rm out}=150$ au, respectively, assuming a power law surface density profile $\Sigma \propto R^{-1}$ at the beginning of the calculation. We model the gas disc with $10^6$ SPH particles assuming a total gas mass of $0.01~M_\odot$ --- see \cite{Price+2017} for details of the disc setup used in {\sc phantom}. We adopt a mean Shakura-Sunyaev disc viscosity $\alpha_{\rm SS} \approx 0.005$ by setting a fixed artificial viscosity parameter $\alpha_{\rm AV} = 0.25$ in the code and using the `disc viscosity' flag (cf. \citealt{Lodato&Price2010}). Furthermore, given the disc mass and temperature considered, we neglect disc self-gravity \citep{toomre64a}. We use a locally isothermal equation of state where the temperature is a function of (spherical) radius from the primary star according to $T(r) = 64 \,{\rm K} \,(r/r_{\rm in})^{-1/2}$. This corresponds to $H/R = 0.05$ at $R = R_{\rm in}$ and $H/R = 0.1$ at $R = R_{\rm out}$. The spherical coordinate choice (instead of cylindrical) is justified because the disc is expected to become twisted during the flyby \citep{Lodato&Price2010}. The disc is always initialized with an inclination equal to $0\degree$ (i.e. with the disc midplane lying in the $z=0$ plane), which will be considered as the reference plane throughout this work.

\subsubsection{Properties of the perturber}

We place the perturber on a parabolic orbit, arriving from the negative $y$ direction and leaving towards negative $y$. We fix the periastron of its orbit to be at $x=0$ and $y>0$ (cf. Fig.~\ref{fig:sketch}). We could choose to incline the orbit in two ways: either about the $y$-axis, which does not affect the three-dimensional periastron position; or about the $x$-axis, which sends the periastron out of the $xy$-plane. Given the axisymmetry of the disc, rotation about the $z$-axis is irrelevant. For simplicity we consider only rotations about the $y$-axis by a (roll) angle, $\beta$ (cf. Fig.~\ref{fig:sketch}). Hence, in our simulations, for a fixed value of periastron distance from the star, $r_{\rm peri}$, all the orbits share the same three-dimensional periastron position. As in \cite{Breslau+2014}, we consider periastron distances ($r_{\rm peri}$) in a range where the disc is perturbed significantly without being fully destroyed ($100 \leq r_{\rm peri} \leq 300$ au).

We vary the inclination of the orbit between $0\degree$ and $180\degree$ in steps of $45\degree$. For angles between $0\degree$ and $90\degree$ the orbit is prograde with respect to the disc rotation, while for angles between $90\degree$ and $180\degree$ it is retrograde. If $\beta=90\degree$, the orbit is polar. Since the goal is to explore the strongest tidal effect for each orbit, we set the eccentricity equal to unity. Parabolic encounters ($e=1$) have the longest interaction time compared to hyperbolic ones ($e>1$) and hence produce the most prominent dynamical signatures.  To avoid artificial effects due to the sudden introduction of a perturber in the field, we set the initial distance of the perturber to $10$ times the periastron distance. This ensures that the perturber's gravitational force onto the closest disc particle is approximately 1\% of the primary's gravitational force. To fully constrain the orbit, we specify the inclination $\beta$ and the periastron distance $r_{\rm peri}$.

We refer to $M_1$ and $M_2$ as the host and the perturber masses, respectively; $M_{\rm t}=M_1 + M_2$ the total mass and $q$ the perturber-to-host mass ratio $M_2/M_1$. In an environment such as the Orion Nebula Cluster, the values of $q$ range between $0.08$ and $500$ \citep{Pfalzner+2005b, Steinhausen+2012}. To keep the study manageable, we assume $M_1=1\,M_\odot$ and consider only the most likely values of $q$: $0.2$, $0.5$, $1$, $2$ and $5$ \citep{Duchene&Kraus2013}. Our flyby setup is publicly available in the {\sc phantom} repository\footnote{ https://bitbucket.org/danielprice/phantom}. Refer to Appendix~\ref{sec:flybysetup} for more detail about the initial setup (positions, velocities, time of flight).

We model both the star and the perturber as sink particles that interact with gas and dust particles via gravity and accretion \citep{Bate+1995,Price+2017}. We set the sink particle accretion radius to $10$ au on the host star and $1$ au on the perturber, with both gas and dust particles accreted provided the usual conditions are met \citep{Price+2017}, preventing the accretion of unbound particles. We accrete particles unconditionally if they pass within a radius less than 80\% of the accretion radius. 

We performed a set of simulations (cf. Table~\ref{tab:sims}) for which we solely changed the perturber's orbital parameters, assuming a single model for the disc around the primary.  For simplicity, we neglect the presence of a disc orbiting around the perturber prior to the encounter, namely the circumsecondary disc. By doing so, we neglect any incoming material that could either be captured or spread around. Hence, assuming that the mass of the potential circumsecondary disc is low, the circumprimary is (almost) unaffected by it. The temperature changes in the circumprimary disc induced by the companion's radiation field are beyond the scope of this paper. Radiative effects during the encounter might be important in generating peculiar observational signatures though, but we defer their study to a follow up paper. Moreover, we stress that the disc that in some cases forms around the perturber is composed of gravitationally bound particles whose temperature is not correctly defined. This does not affect the evolution of the circumprimary disc.

\subsection{Dust modelling}
\label{sec:dustmethod}

We set up the dust disc to follow the same radial density profile as the gas disc at the beginning of the simulation, but with the dust mass scaled down by a factor of $100$ from the gas mass. 
We model the dust grain sizes ranging from $1$~$\mu$m to $10$~cm. This allows us to explore a wide range of aerodynamical drag regimes, from tightly-coupled particles (with size $1$~$\mu$m) to poorly-coupled particles (with size $10$~cm). In the {\sc phantom} code, dust dynamics can be computed using two different methods:  with dust and gas modelled as separate sets of particles (`two fluid') \citep{Laibe&Price2012, Laibe&Price2012b} or using one set of particles to represent the mixture (`one fluid') \citep{price15a,ballabio18a}. The latter is optimized to simulate particles tightly-coupled to the gas in the terminal velocity approximation \citep[e.g.][]{youdin05a}. 
Which regime is appropriate depends on the Stokes number --- the ratio of the orbital time-scale to the drag stopping time $t_{\rm s}$ (the time-scale for the drag to damp the local differential velocity between the gas and dust). In the linear Epstein regime \citep{epstein24a}, the Stokes number is given by
\begin{equation}
{\rm St} \equiv t_{\rm s} \, \Omega_{\rm K} = \frac{\rho_{\rm gr} s}{\rho c_{\rm s}} \, \Omega_{\rm K}  ,
\label{eq:stokes}
\end{equation}
where $s$ is the grain size, $\Omega_{\rm K}$ is the Keplerian angular speed, $c_{\rm s}$ the speed of sound, $\rho_{\rm gr}$ is the bulk grain density and $\rho$ is the total density. If we assume $\rho_{\rm gr}=3 \,\, {\rm g \, cm^{-3}}$, given the disc model considered, then the particles of $\sim 1$ mm have ${\rm St}\approx 1$ in the outer disc regions. The dust distribution and total mass in the dust populations in our sample is the same, regardless of the grain size. Therefore, we underestimate the effect of the cumulative back reaction \citep{hutchison18a,dipierro18b}, which is expected to be negligible given the 1\% dust-to-gas considered in this study.

The aerodynamical coupling between the gas and dust changes with time due to the rapidly changing gas density during the flyby. This means that grains that are well-coupled at the beginning of the simulations can become de-coupled during and after the encounter. We therefore performed a set of tests to determine the best method for each grain size, based on the evolution of the Stokes number \citep{Laibe&Price2012,price15a}. We found that the one-fluid method was most suitable for $s=[1,10]$ $\mu$m, while the two-fluid method was more efficient for $s=[0.1,1,10,100]$ mm (cf. Appendix~\ref{sec:1vs2fluid}). When using the one-fluid method, we model the circumprimary disc as a fluid made of a mixture of $10^{6}$ particles; while in our two-fluid simulations we consider a disc made of $10^{6}$ gas particles and $4 \times 10^{5}$ dust particles, respectively.

\subsection{Simulation set}
\label{sec:simsset}
We performed a set of simulations for which we solely changed the perturber's orbital parameters, assuming a single model for the disc around the primary.
Table~\ref{tab:sims} lists our simulation ensemble, broken into subsets. In each subset, we examine how the dynamics depends on a particular orbital parameter of the perturber. We first considered the case of an unperturbed disc --- without any stellar perturber --- which can be considered as the reference case (NoFB). We then varied the inclination of the perturber's orbit from $0\degree$ to $180\degree$ in steps of $45\degree$ at fixed mass ratio and periastron separation. For the most interesting cases, namely the inclined prograde and retrograde orbits, we considered different mass ratios $q$ ($0.2, 0.5, 1, 2, 5$) and values of the periastron distance ($100$, $200$ and $300$ au). The results of simulations with different $q$ are reported in Appendix~\ref{sec:appendix-kin-q}. 

\begin{table}
\begin{center}
\caption{Set of flyby simulations for different perturber's orbital parameters (periastron, mass ratio and inclination) and dust sizes ($s$).}
\label{tab:sims}
\begin{tabular}{|c|c|c|c|c|c|}
\hline
Name & $r_{\rm peri}$ (au) & $q$ & $\beta \, (\degree)$ & $s$ \\
\hline
\hline
    NoFB & $\infty$ & 0 & 0 & 1 cm \\
\hline
    $\beta$0 & 200 & 1.0 & 0 & 1 cm \\
	$\beta$45 & 200 & 1.0 & 45 & all \\
	$\beta$90 & 200 & 1.0 & 90 & 1 cm \\
	$\beta$135 & 200 & 1.0 & 135 & all  \\
	$\beta$180 & 200 & 1.0 & 180 & 1 cm \\
\hline
    $\beta$45-$r_{\rm p}$100 & 100 & 1.0 & 45 & 1 cm \\
    $\beta$45-$r_{\rm p}$300 & 300 & 1.0 & 45 & 1 cm \\
    $\beta$135-$r_{\rm p}$100 & 100 & 1.0 & 135 & 1 cm \\
    $\beta$135-$r_{\rm p}$300 & 300 & 1.0 & 135 & 1 cm \\
\hline
    $\beta$45-q02 & 200 & 0.2 & 45 & 1 cm \\
    $\beta$45-q05 & 200 & 0.5 & 45 & 1 cm \\
    $\beta$45-q2 & 200 & 2 & 45 & 1 cm \\
    $\beta$45-q5 & 200 & 5 & 45 & 1 cm \\
\hline
    $\beta$135-q02 & 200 & 0.2 & 135 & 1 cm \\
    $\beta$135-q05 & 200 & 0.5 & 135 & 1 cm \\
    $\beta$135-q2 & 200 & 2 & 135 & 1 cm \\
    $\beta$135-q5 & 200 & 5 & 135 & 1 cm \\
\hline
    $\beta$45-nd & 200 & 1.0 & 45 & no drag \\
    $\beta$135-nd &200 & 1.0 & 135 & no drag \\
\hline
\end{tabular}
\end{center}
\end{table}

We performed a complete set of simulations covering the full range of grain sizes only for $\beta$45 and $\beta$135 since the dust-gas calculations are computationally expensive. We also performed a two-fluid simulation where we switched off the drag interaction between the gas and dust phases, to investigate the response of test particles to the tidal encounter (`no-drag'). In this case, the dynamics is similar to previous studies \citep{Breslau+2014,Bhandare+2016,Breslau+2017}. In all the other cases shown in Table~\ref{tab:sims}, we modelled only cm-sized particles.

\begin{figure*}
\begin{center}
\includegraphics[width=0.92\textwidth]{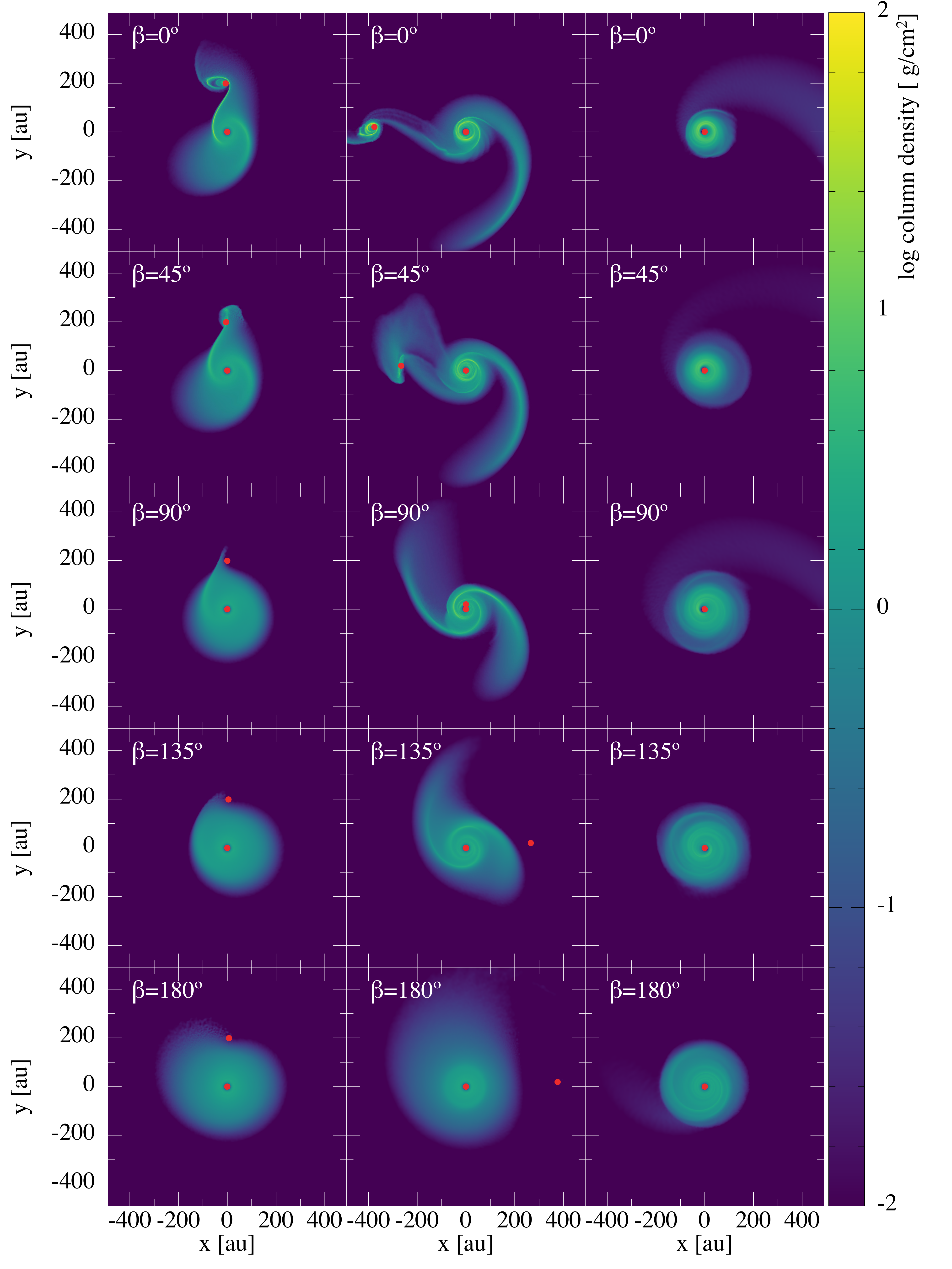}
\caption{Face-on view of the gas column density for (from top to bottom rows): $\beta0$, $\beta45$, $\beta90$, $\beta135$, $\beta180$. From left to right columns in each panel: t=5\,400 yr (periastron), t=5\,950 yr, t=8\,100 yr. The disc rotation is anticlockwise. Sink particles (in red) are large for visualization purposes only. Spirals are clearly seen along most of the perturber's close approach to the disc, except for $\beta$180. A bridge appears between both stars shortly after the periastron for prograde orbits. The size of the disc increases with $\beta$: prograde encounters truncate the disc more heavily compared to retrograde ones.}
\label{fig:gas-panel}
\end{center}
\end{figure*}

\begin{figure*}
\begin{center}
\includegraphics[width=0.92\textwidth]{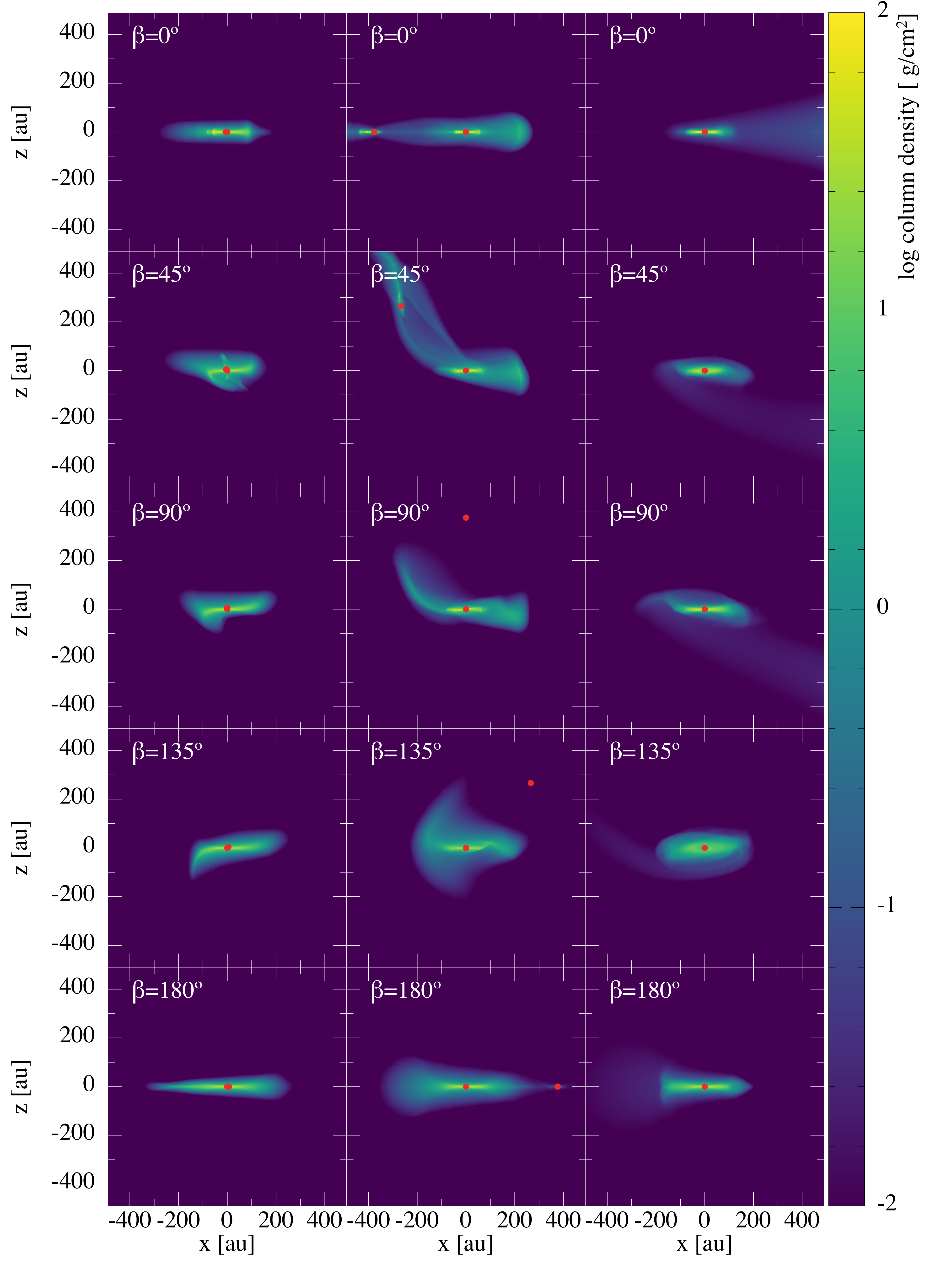}
\caption{Edge-on view of the gas column density for $\beta0$, $\beta45$, $\beta90$, $\beta135$, $\beta180$ (top to bottom). Left to right columns show t=5\,400~yr (periastron), t=5\,950 yr, and t=8\,100 yr. Sink particles (in red) are large for visualization purposes only. Perturbers on non-coplanar orbits lift and push material out of the initial ($z=0$) disc plane. Shortly after the periastron, the disc is warped. After 2\,700 yr, the remaining disc is slightly tilted for inclined orbits.}
\label{fig:gas-panelz}
\end{center}
\end{figure*}

\section{Results}
\label{sec:results}

\subsection{Gas density evolution}
\label{sec:gasdensity}

\subsubsection{Prograde vs. retrograde}
Figures~\ref{fig:gas-panel} and \ref{fig:gas-panelz} show the column density of the gas for $\beta$0, $\beta$45, $\beta$90, $\beta$135 and $\beta$180 (top to bottom rows, respectively) as a function of time during the flyby. Figure~\ref{fig:gas-panel} represents the $xy$-view, while Figure~\ref{fig:gas-panelz} shows the $xz$-view. For each case, we show three evolutionary stages (from left to right): 5\,400~yr (periastron), 5\,950~yr and 8\,100~yr, respectively. As expected, we observe the development of prominent two-armed spiral structures, in which one of the arm is sometimes significantly stronger than the other. We find results for the coplanar configurations ($\beta$0 and $\beta$180) in agreement with \cite{Clarke&Pringle1993}: a prograde orbit is more destructive and leads to more prominent spiral arms compared to a retrograde encounter. The perturber on a prograde orbit strips the disc and captures a significant amount of material. Intermediate orbital inclinations ($45\degree \leq \beta \leq 135\degree$) produce a range of structures that smoothly connect the results of our prograde coplanar and retrograde coplanar simulations.

\begin{figure*}
\begin{center}
\includegraphics[width=0.92\textwidth]{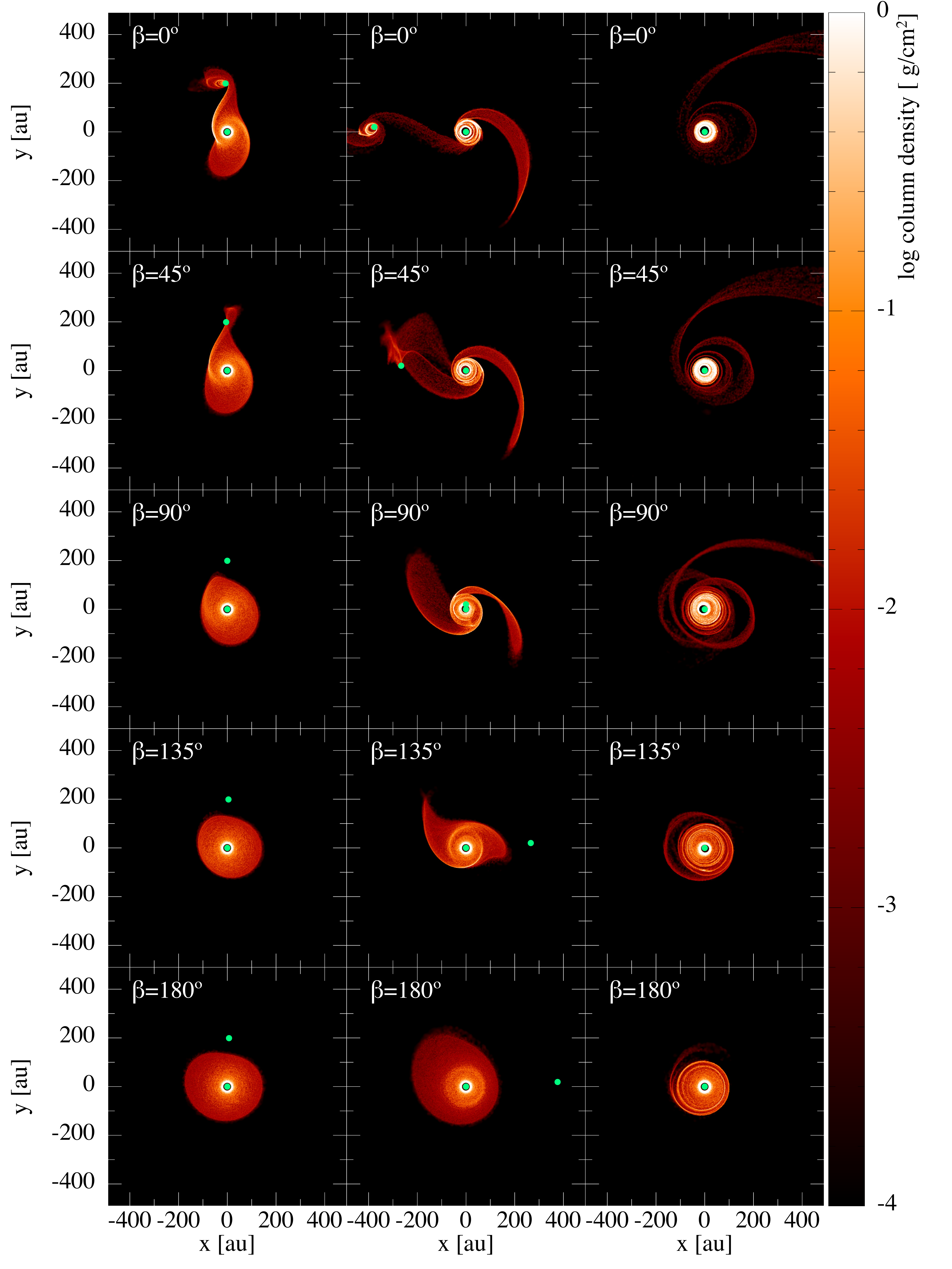}
\caption{Face-on view of the dust column density for (top to bottom): $\beta0$, $\beta45$, $\beta90$, $\beta135$, $\beta180$. From left to right columns in each panel: t=5\,400 yr (periastron), t=5\,950 yr, t=8\,100 yr. The disc rotation is anticlockwise. Sink particles (in green) are large for visualization purposes only. Similar features as in Fig.~\ref{fig:gas-panel} are observed: spirals, bridges and truncated discs. The radial extent of the disc when the encounter occurs is smaller than the gas, because of radial drift. The structures are sharper both due to gravitational and drag effects (cf. Sect.~\ref{sec:geomdrag}).}
\label{fig:dust-panel}
\end{center}
\end{figure*}

\begin{figure*}
\begin{center}
\includegraphics[width=0.92\textwidth]{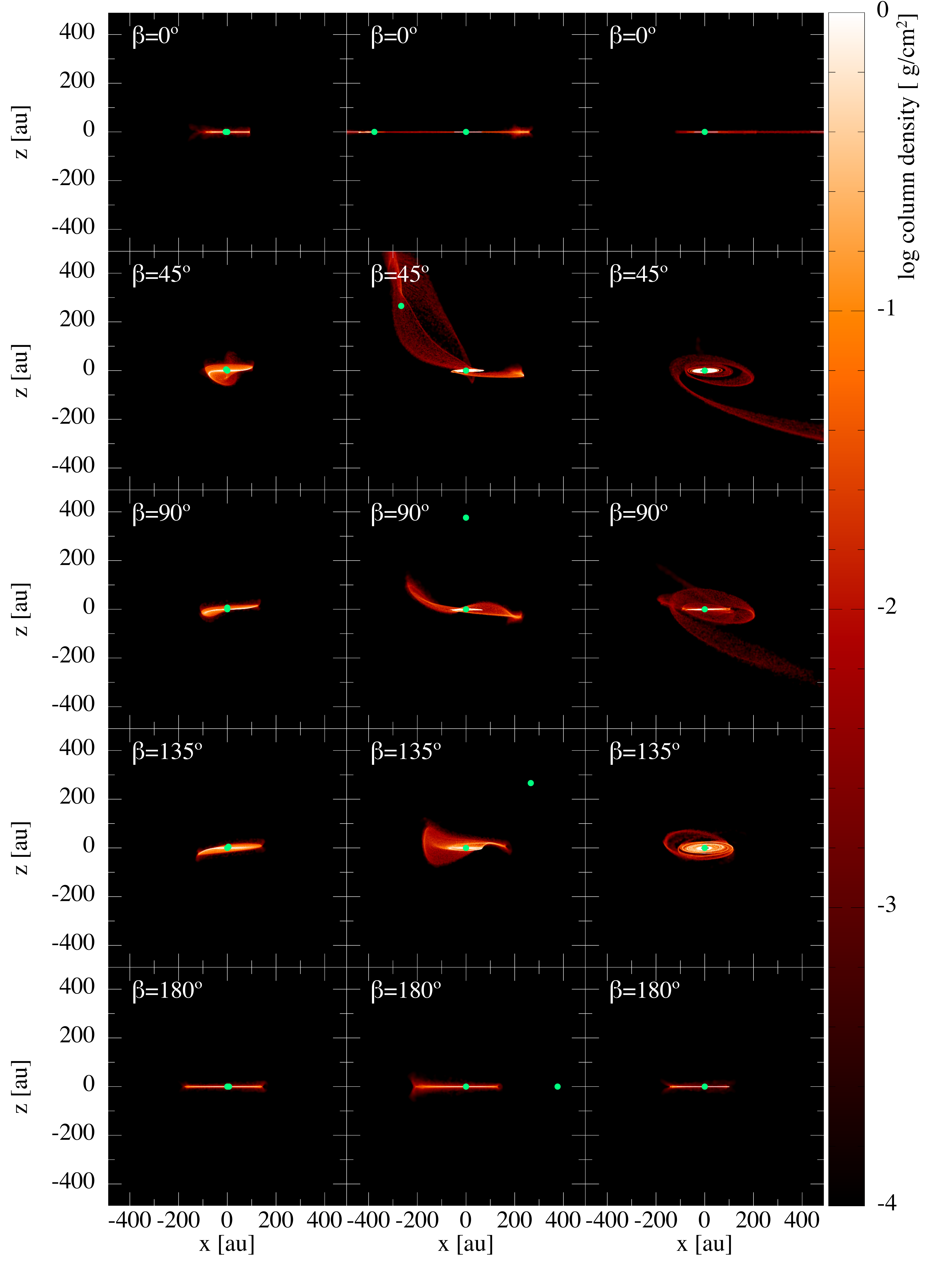}
\caption{Edge-on view of the dust column density for (top to bottom): $\beta0$, $\beta45$, $\beta90$, $\beta135$, $\beta180$. From left to right columns in each panel: t=5\,400 yr (periastron), t=5\,950 yr, t=8\,100 yr. Sink particles (in green) are large for visualization purposes only. Similar features as in Fig.~\ref{fig:gas-panelz} are observed: material out of the initial ($z=0$) disc plane and warps. The dusty discs are very thin in the vertical direction because of drift effects. This leads to efficient dust settling towards the disc mid-plane. For perturbers on inclined orbits, the remaining disc after 2\,700 yr is slightly tilted.}
\label{fig:dust-panelz}
\end{center}
\end{figure*}

\subsubsection{Spirals and bridges}

When the perturber is at periastron (left columns in Figs.~\ref{fig:gas-panel} and \ref{fig:gas-panelz}), retrograde and polar orbits are hard to distinguish, whereas in the prograde case the disc already exhibits an asymmetric morphology and a sharp spiral or ``bridge'' that connects both stars. Shortly after the passage at the periastron, each orbital configuration has a more distinctive signature in the density field. As expected, perturbers on prograde and polar orbits trigger the formation of spirals in the disc. However, spirals are also observed in the \textit{inclined retrograde} case ($\beta$135). The main differences between polar and prograde orbits are i) the amount of material captured by the secondary star and ii) the morphology of the spiral structure. Hereafter, whenever we refer to the sharpness of the spiral structure we mean the width of the spiral arms. For instance, using the definition above, the spiral on the right-side of the disc is sharper in $\beta$45 compared to the one in $\beta$135. Once the perturber has left the field, the main noticeable differences are the kinematics of the disc (discussed in detail in Section~\ref{sec:kinematics}, below) and its size (discussed in Section~\ref{sec:truncation}, below).

\subsubsection{Disc warping and diffuse halo}

The $xz$-snapshots of Fig.~\ref{fig:gas-panelz} show that during and after the encounter disc material is lifted out of the $z=0$ plane. This is particularly striking for inclined orbits where the spirals on each side of the disc are out of the plane as well. For $\beta$45 and $\beta$135, the inclination and the orientation of the disc are modified during the encounter (cf. Sect.~\ref{sec:warps}). Hence, the disc is warped by the stellar perturber. By contrast, we notice that the disc plane remains almost unperturbed for coplanar and polar orbits, in agreement with with \cite{XG2016}. Remarkably, for $\beta$45 a ``bridge'' of material connects both stars $\sim500$ yr after the periastron passage. In all our simulations, when the perturber leaves the field of view of Fig.~\ref{fig:gas-panel} (i.e. at a distance of more than 500 au) there is a diffuse halo of gas around the circumprimary disc, caused solely by the stellar flyby.

\subsection{Dust density evolution}
\label{sec:dustdensity}

\subsubsection{Prograde vs. retrograde}

Fig.~\ref{fig:dust-panel} shows the corresponding $xy-$view of the dust column density (in centimetre-sized grains). Snapshots are the same as in Fig.~\ref{fig:gas-panel} but showing the dust instead of the gas. As in the gas evolution, the dust disc is more heavily affected if the perturber is on a prograde orbit compared to a retrograde orbit. This may be seen by comparing our results with $\beta=0\degr$ (top row) to $\beta=180\degr$ (bottom row).

\subsubsection{Spirals and bridges}

The tidally-induced structures in the disc are sharper and more well defined in the dust compared to the gas, both radially and azimuthally. This can be explained by the difference in the relative size of the dust disc compared to the gas disc \emph{before} the encounter. Because large dust grains undergo radial-drift \citep{W77, Laibe2012}, the outer edge of the dust disc prior to the encounter is smaller than the gas disc. Therefore, the closest approach between the perturber and the gas and dust discs is different (see Section~\ref{sect:dependgrainsize}, below).

\begin{figure*}
\begin{center}
\includegraphics[width=\textwidth,trim={1.2cm 0.6cm 0 0},clip]{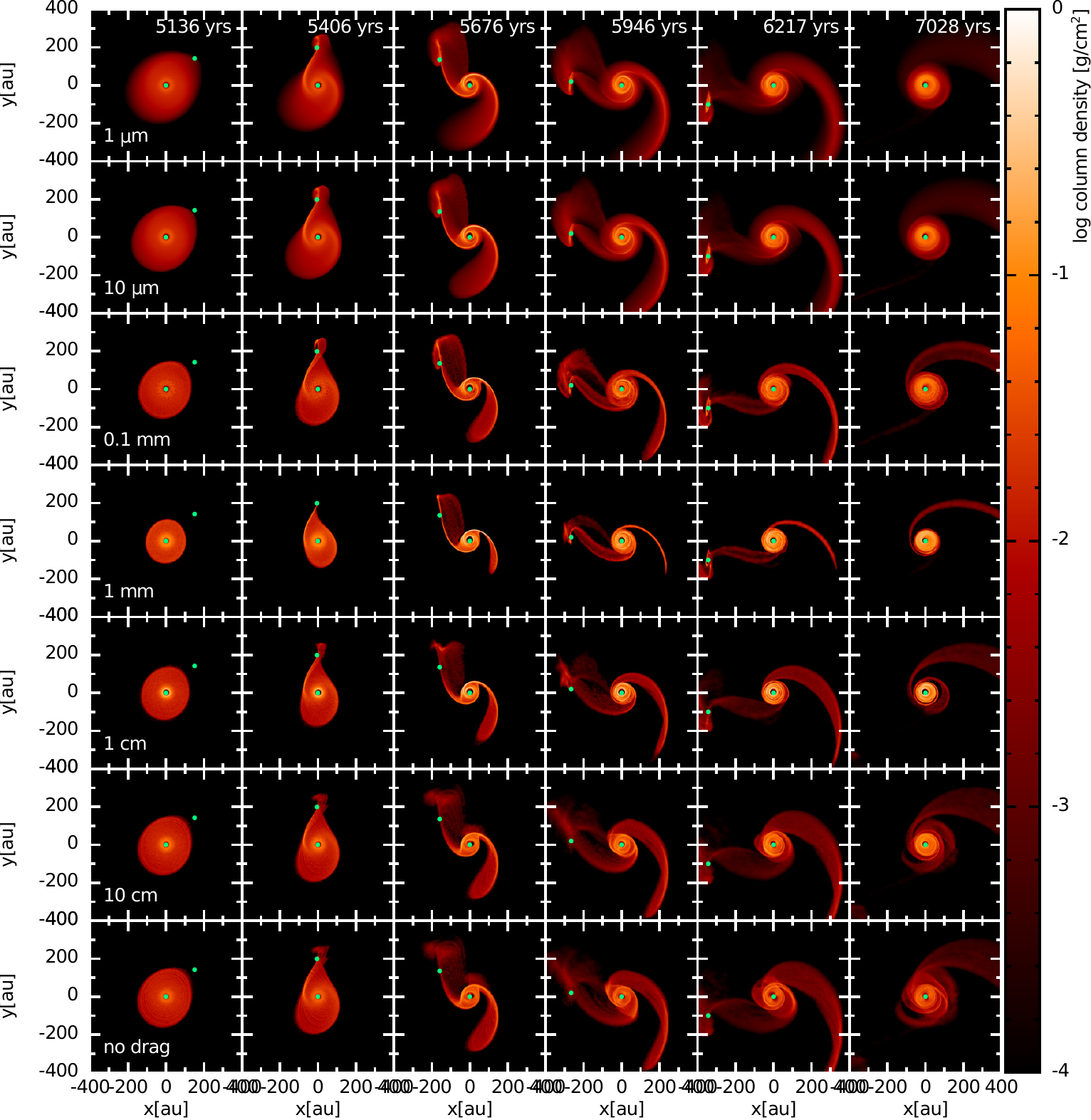}
\caption{Face-on views of the dust disc for different grain sizes (top to bottom) and as a function of time (left to right) in the disc model $\beta$45. The last row shows the dust evolution in absence of the gas. Each panel has 400 au in width and in height. The disc rotation is anticlockwise. Sink particles (in green) are large for visualization purposes only. At the periastron, the 1 mm dust disc is more compact than the 1 $\mu$m one due to radial drift. The resulting spirals in the mm are sharper, i.e. narrower, compared to other grain sizes because of gas drag. Micron-sized particles behave as gas particles given that they are strongly-coupled to the gas; while 10 cm particles, poorly-coupled to the gas, behave as N-body test particles (no drag).}
\label{fig:dustr45}
\end{center}
\end{figure*}

\begin{figure*}
\begin{center}
\includegraphics[width=\textwidth,trim={1.2cm 0.6cm 0 0},clip]{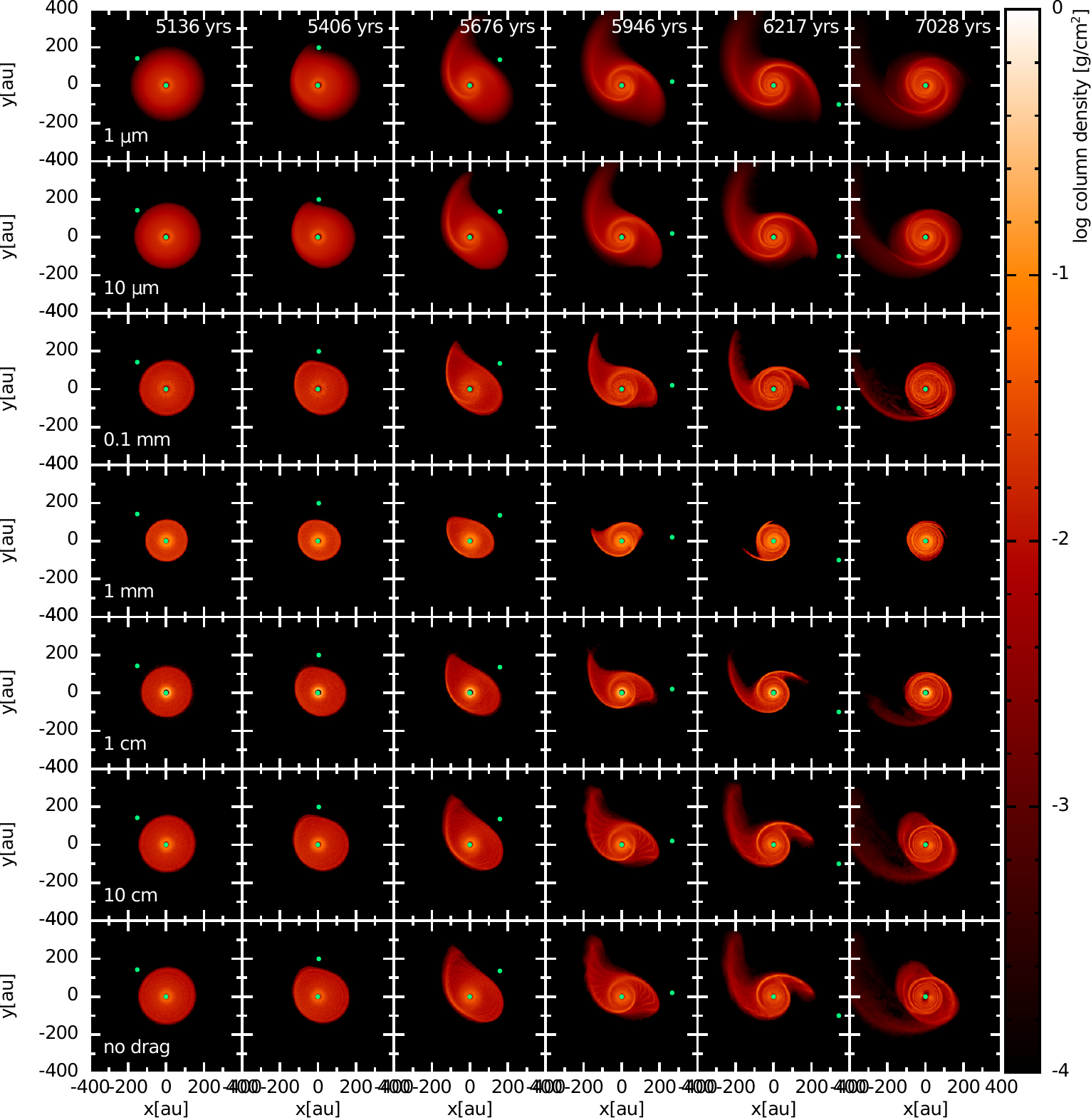}
\caption{Face-on views of the dust discs for (from top to bottom) different grain sizes and (from left to right) as a function of time in the disc model $\beta$135. The last row shows the dust evolution in absence of the gas. Each panel has 400 au in width and in height. The disc rotation is anticlockwise. Sink particles (in green) are large for visualization purposes only. The remarks of Fig.~\ref{fig:dustr45} are also valid for this case. The main difference is that in the inclined retrograde case less prominent spirals form in the dust.}
\label{fig:dustr135}
\end{center}
\end{figure*}

\subsubsection{Dependence on grain size}
\label{sect:dependgrainsize}
Figures~\ref{fig:dustr45} and \ref{fig:dustr135} show the response of grains with different sizes to the flyby. We show the results of a series of calculations with grain sizes $s=[10^{-3}, 10^{-2}, 0.1, 1, 10, 100]$ mm (top to bottom, respectively) for the two most interesting and likely cases: $\beta$45 (Fig.~\ref{fig:dustr45}) and $\beta$135 (Fig.~\ref{fig:dustr135}). Assuming that tidal encounters are stochastic, inclined orbits (either prograde or retrograde) are more likely than coplanar or polar configurations. 

Micron-sized dust grains, as expected, closely follow the gas density, exhibiting similar prominent spiral structures. For prograde orbits, a significant fraction of these small dust grains are captured into a circumsecondary disc by the perturber. 

The mm grains, by contrast, form a more compact disc with sharper (i.e., radially narrower) spiral arms. We adopt the same colourbar in Figs.~\ref{fig:dustr45} and~\ref{fig:dustr135} for the different grain sizes. This allows a direct comparison between the images. These grains are only marginally captured into the circumsecondary disc. In addition, close to the star, radial drift causes the dust-to-gas ratio to increase for mm and cm-sized grains (see e.g. Fig.~\ref{fig:sigmavsr}). 

For larger grain sizes ($s=[1, 10]$ cm), we recover a more prominent and extended spiral structure. In the limit of grains with size equal to $10$ cm, almost unaffected by radial-drift, dust particles behave as test particles in a gas-free environment. To check this, we also performed a simulation with no drag (i.e. test particles) to be compared with the $10$-cm dust distribution. The two lower rows of Figs.~\ref{fig:dustr45} and \ref{fig:dustr135} show almost identical distributions in the dust for these two particular cases, which confirms our hypothesis.

Our simulations thus predict that tidally-induced spirals should be more readily observed in scattered light compared to dust thermal emission, especially for retrograde orbits (compare the micron and mm dust surface density in Fig.~\ref{fig:dustr135}). Scattered light traces micron-sized particles while the dust thermal emission is sensitive to mm grains. Large differences between scattered light and long-wavelength dust continuum images have been revealed in a number of sources, which might be explained by this mechanism (see Sect.~\ref{sec:flybymechanism}).

\subsection{What physical effects dominate the dust evolution?}
\label{sec:geomdrag}

What is the primary reason for the sharper spiral structure in the ${\rm St} \approx 1$ dust grains? We investigate four hypotheses: i) the smaller dust disc size from radial drift \emph{prior} to the encounter ii) trapping of dust particles in the pressure gradients induced by spiral arms iii) drag-induced circularization of the dust orbital motion \citep{adachi76a} or iv) the absence of viscous and pressure forces.

To distinguish between these hypotheses, we performed an additional set of simulations of micron-sized dust particles, adopting smaller values of the outer disc radius compared to the reference case ($R_{\mathrm{out}}=80$ and $100$ au, rather than $150$ au) in order to reproduce the geometry of the mm dust disc just before the flyby (see middle row in Fig.~\ref{fig:dustr45}). We focus our analysis on the inclined prograde case ($\beta45$), since the interaction is more effective in shaping the resulting structure (see Fig.~\ref{fig:dustr45}). For these test simulations, we scaled the disc mass in each case so that 0.01 $M_{\odot}$ is always distributed between $10$ and $150$ au in order to guarantee the same Stokes number for micron sized grains as in the reference simulations (cf. Eq.~\ref{eq:stokes}).

For a meaningful comparison, we need to hold fixed both the local gas density and the Stokes number. Therefore, we choose the same conditions for the gas and dust discs for all the disc extent. Doing otherwise, e.g. only changing the dust disc extension, would create a significant gradient in the dust-to-gas ratio at the outer edge of the dust disc. Then, the region beyond that limit would not be comparable to the reference simulations, preventing us from identifying the underlying physical mechanism.

In order to assess the influence of drag in trapping dust grains (hypothesis ii) and in recircularizing the orbit of dust particles (hypothesis iii), we also performed the same set of simulations with the drag switched off (`no-drag').

\subsubsection{Dust disc size prior to the encounter}
\label{sec:dustdiscsize}

Comparison of the leftmost panels of Figs.~\ref{fig:dustr45} and \ref{fig:dustr135} suggests that the primary reason for the different morphology of the spiral structure is the disc geometry preceding the flyby.  This discrepancy is observed in nature: isolated discs show evidence for radial size-sorting of grains \citep[e.g.][]{tazzari15a}, induced by the combination of radial drift and grain growth (see the review by \citealt{testi14a} and references therein).

Figure~\ref{fig:dustr45rout} shows that, as expected, a flyby of a disc with a smaller outer radius produces a less extended and weaker spiral structure in both gas and small dust grains. Comparing the morphology of the spiral structure of mm-sized dust grains in Fig.~\ref{fig:dustr45} with the resulting spiral structure of micron-sized grains in a disc with outer radius $R_{\mathrm{out}}=80$ au (the simulation that best resembles the extent of the mm dust disc just before the flyby), the smaller radial extent of the dust disc induced by the drag \emph{before} the flyby is the main factor affecting the post-flyby spiral structure.

\subsubsection{Dust trapping in spiral arms}

While dust trapping induced by the presence of pressure maxima in the radial disc structure has been thoroughly studied \citep[e.g.][]{Laibe2012}, the influence of gas spiral structure on the dust dynamics has been investigated in only a few cases \citep{Clarke&Lodato2009, Pinilla+2012,dipierro15a,Cuello+2018}. The classical picture is that particle trapping in non-axisymmetric overdensity regions is only possible for structures which are close to corotation with the local gas flow, like spiral arms in gravitationally unstable discs \citep[e.g.][]{booth16a}. By contrast, spirals induced by an embedded planet are not expected to be able to trap dust particles (see \citealt{Ayliffe+2012}) since they corotate with the planet, thus moving with respect to the background gas flow. More generally, spiral arms are able to successfully concentrate dust if, and only if, the characteristic lifetime for the spiral features is larger than the concentration timescale.

\begin{figure*}
\begin{center}
\includegraphics[width=\textwidth,trim={1.3cm 0.7cm 0 0},clip]{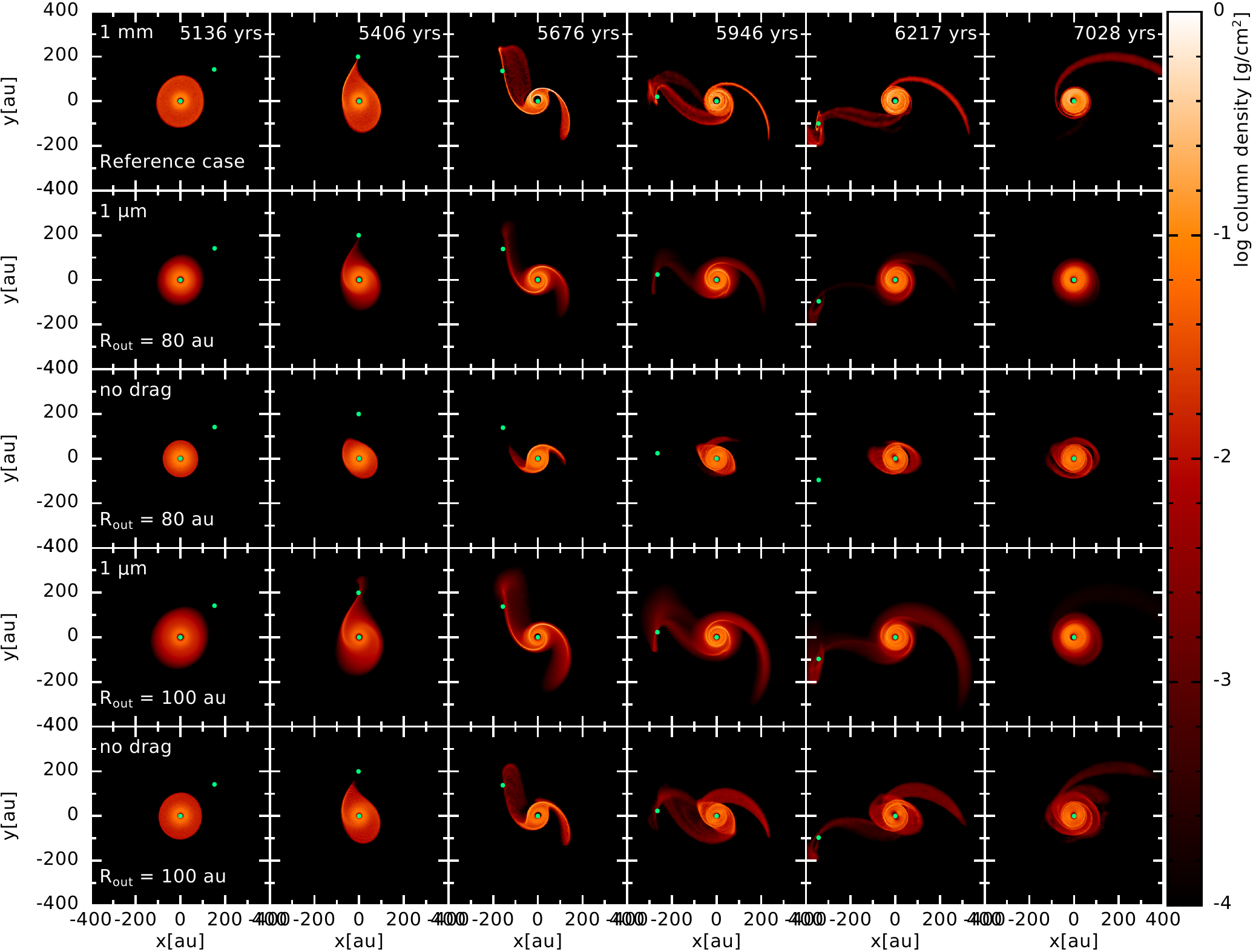}
\caption{Face-on views of the dust discs with different outer radius for micron-sized grains (second and forth rows) and 'no drag' case (third and fifth rows) along the flyby in the disc model $\beta$45 (Fig.~\ref{fig:dustr45}), compared to the mm-sized dust density evolution of the reference case (upper panels). Each panel has 400 au in width and in height. The disc rotation is anticlockwise. Sink particles (in green) are large for visualization purposes only. The 1 mm reference case (middle row of Fig.~\ref{fig:dustr45}) is shown again in the top panels for convenience.
The trapping of dust in the spirals is due to both the geometrical effect of the flyby and the gas drag. This can be seen by comparing the 1 mm reference case to the 1 $\mu$m and ``no drag'' cases, for different values of $R_{\rm out}$.}
\label{fig:dustr45rout}
\end{center}
\end{figure*}

From the two upper panels of Fig.~\ref{fig:dustr45rout} we see that mm-sized grains are more concentrated onto the spiral structure compared to micron-sized grains in the disc model with $R_{\mathrm{out}}=80$~au, supporting the idea that the flyby-induced spiral structure can trap dust. Comparing the morphology of the spiral structure of mm dust grains (the top panels of in Fig.~\ref{fig:dustr45rout}) with the spiral morphology in the middle row of Fig.~\ref{fig:dustr45rout} (`no-drag' case) corresponding to $R_{\mathrm{out}}=80$~au, we note that when the perturber is at periastron the structure is similar. However, just after the periastron, the spiral structure in mm grains appears more concentrated. This indicates again that the sharpness of the dust features is enhanced by the drag. In other words, the drag renders the spirals narrower. Assuming that the gas spiral structure is corotating with the local gas flow, we can simply model the spiral structure as a collection of azimuthally limited pressure bumps in the radial direction moving with Keplerian velocity.

The timescale to accumulate dust particles in the pressure maxima is expected to be shorter than the global drift time-scale. In detail, if the pressure inhomogeneity has a radial scale length $\Delta R$, the local radial dust velocity is $\sim \Delta v \propto \partial P/\partial R \sim P/\Delta R$ \citep[e.g.][]{Laibe2012}, exceeding the unperturbed radial dust motion $\propto P/R$. The time-scale for solids to pile-up at the pressure maximum is further reduced by factor $\Delta R/R$ because the large grains have to move only a distance of order $\Delta R$ to reach the density maxima. If we assume that these density and pressure inhomogeneities are characterized by a typical length scale $\sim H_{\mathrm{g}}$, the time-scale to concentrate solids at the pressure maxima is shorter than the global drift time-scale by a factor of $\left(\Delta R/R\right)^2\sim\left (H_{\mathrm{g}}/R \right)^2$, approaching  values of the order of $\Omega_{\mathrm{K}}^{-1}$ for marginally-coupled particles in typical discs. This estimate is consistent with the dust concentration efficiency inside spiral arms in self-gravitating discs derived by \cite{Clarke&Lodato2009}.

Interestingly, we also see that the dust efficiently accumulates where the amplitude of the spirals is the strongest, i.e. at large stellocentric distances. This mechanism can be interpreted as a trajectory crossing effect \citep{Clarke&Pringle1993}: because of the drag, the relative velocities of particles with St $\sim 1$ are quickly damped when they pass through the spiral. If $t_{\rm s} \ll \Omega^{-1}$ the dust stops before it reaches the maximum in the spiral, while for $t_{\rm s} \gg \Omega^{-1}$ the dust passes through the spiral. This is in agreement with our results where mm particles, with St $\sim 1$ at the spiral location, are more concentrated than the other grain sizes.

\subsubsection{Drag-induced circularization of the dust disc}

Figure~\ref{fig:dustr45rout} allows us to compare how the dust disc response depends on the coupling to the gas. This is done for dust discs with similar radial extensions ($80-100$ au) when the perturber is at pericentre. After the encounter, we observe that the mm dust particles (top row) quickly circularize around the star, while the decoupled particles (`no-drag') still move on elliptical orbits. This is related to the circularization effect induced by the drag on dust particles, which tends to damp the eccentricity of dust particles on a time-scale of the order of $\left (\mathrm{St}+\mathrm{St}^{-1}\right) \Omega_{\mathrm{K}}^{-1}$ \citep{adachi76a}. This effect explains why the disc structure after the flyby is more radially concentrated for the mm dust particles compared to the `no-drag' case.

\subsubsection{Physical effects affecting dust evolution}

The flyby-induced morphology strongly depends on the disc size prior to the encounter. For instance, more compact dust discs --- due to radial-drift --- produce less extended spiral structures. Interestingly, the gaseous spirals are able to efficiently trap dust. Finally, the gas drag has a double effect on dust particles: it renders the dust spirals sharper, and it causes the quick circularization of dust particles around the star. These effects are particularly relevant for dust particles with St~$\sim$~1: mm- to cm-sized particles in this study.

\subsection{Spirals with evolving pitch angles}

The evolving pitch angle of the `grand-design' spirals during the encounter, especially for prograde orbits, is the `smoking gun' of a flyby. We define a grand-design spiral as a well organised large-scale spiral structure in the disc (as opposed to flocculent spirals), as in galactic dynamics and self-gravitating discs. Shortly after periastron passage, the pitch angles of the spirals are approximately $25-30\degree$. Interestingly, the spiral on the side of the disc where the perturber is located has a smaller pitch angle compared to the one on the opposite side. The former is due to the gravitational perturbation itself, while the latter is caused by the movement of the centre of mass \citep{Pfalzner2003}. The measured difference is $\sim 5\degree$. Hence, the pitch angles of the two diametrically-opposed spirals are not equal (cf. Section~\ref{sec:flybymechanism}). These grand-design spirals are transient, evolving into azimuthal asymmetries in the disc once the perturber leaves the field (Figs.~\ref{fig:gas-panel}, \ref{fig:gas-panelz}, \ref{fig:dust-panel} and \ref{fig:dust-panelz}), typically after a few thousand years. At 50 au, where the orbital period is of approximately $350$ yr, this corresponds to several orbits around the central star. These discrepancies between dust species --- in radial extent and sharpness --- represent the dynamical signature of a flyby, past or present (cf. Section~\ref{sec:flybymechanism}).

\subsection{Kinematic signatures}
\label{sec:kinematics}

\begin{figure*}
\begin{center}
\includegraphics[width=\textwidth]{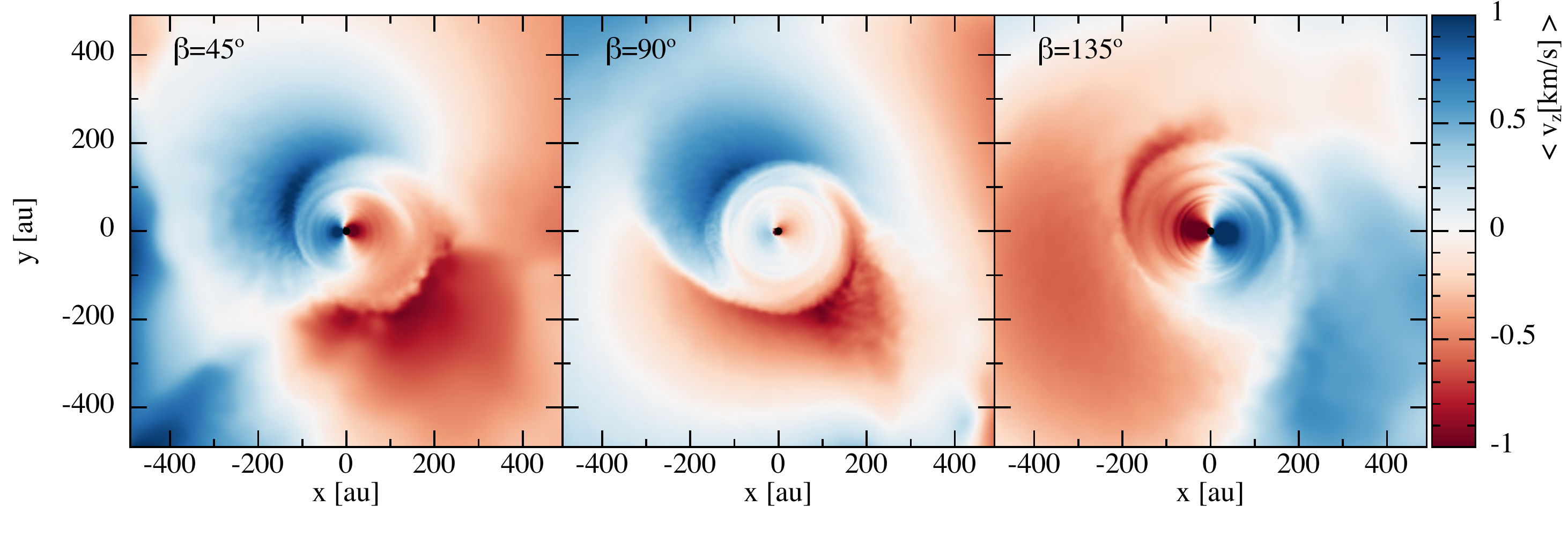}
\caption{Kinematics induced by the flyby, showing mass weighted integral of $v_z$ along the line of sight. From left to right: $\beta45$, $\beta90$ and $\beta135$ after 2\,700 yr after periastron passage. The disc rotation is anticlockwise. The perturber, at a distance of approximately 1200 au, has already left the but the kinematic signatures remain. The location of the red-shifted and blue-shifted regions of the disc provide information about the perturber's orbit inclination. A companion on a polar orbit coming from above and leaving below the disc produces the same vertical velocity departures, but with opposite signs. The unperturbed and coplanar cases produce homogeneous (white) images in $v_z$.}
\label{fig:kinematics}
\end{center}
\end{figure*}

 Figure~\ref{fig:kinematics} shows the vertical (out-of-plane) gas velocity integrated along the line of sight 2\,700 years after the closest approach. Specifically, we show the mass weighted integral
\begin{equation}
\langle v_z \rangle \equiv \frac{\int \rho v_z {\rm d}z}{\int \rho {\rm d} z}.
\end{equation}
We focus on the inclined orbits $\beta$45 (left), $\beta$90 (middle) and $\beta$135 (right) since these are expected to lift and push material out of the disc plane. Coplanar orbits (not shown in the Figure) show only small out-of-plane motions, exhibiting an almost homogeneous vertical velocity field. For $\beta$45, the perturber arrives from below the disc from the right-hand side and leaves from above the disc on the left-hand side of the Figure. This is in agreement with the negative velocities ($\sim -1 \,{\rm km\,s^{-1}}$) seen in the bottom right corner and the positive velocities ($\sim 1 \,{\rm km\,s^{-1}}$) in the top left corner in Fig.~\ref{fig:kinematics}. For $\beta$135, the sign of the velocities is switched because the perturber approaches below the disc from the left-hand side and leaves above the disc on the right-hand side in the reference frame of the Figure. The velocity field in $\beta$90 is intermediate between $\beta$45 and $\beta$135. Here, the perturber arrives from below the disc and goes through periastron in the northern part of the disc.

These kinematic signatures are modestly blurred and smoothed by the disc differential rotation and viscosity. These velocity anomalies can in principle constrain the perturber's orbital inclination, even in the absence of its direct detection. For example, after 2\,700 yr the perturber is already at 1200 au from the primary star, hence potentially out of the observing field. Larger values of $q$ or smaller $r_{\rm peri}$ increase the negative and positive deviations from $0$ in the velocity field (cf. Appendix~\ref{sec:appendix-kin-q}). These kinematic signatures can also help to constrain the disc warping due to the flyby (cf. Sect.~\ref{sec:warps}, below).

An inner binary can generate similar signatures in the disc. See for example figures 9 and 10 in \cite{Juhasz&Facchini2017}. The warped discs presented in that study have negative velocities on one side and positive velocities on the diametrically-opposite side (as in Figure~\ref{fig:kinematics}). This pattern is not seen in unperturbed discs. Interestingly, because of projection effects, the kinematical signatures decrease for increasing disc inclination. In particular, warped kinematics are barely visible in radial velocity for edge-on discs because they are mostly perpendicular to the line of sight.

\subsection{Disc eccentricity pumping by the flyby}

In Figure~\ref{fig:eccentricity} we show the eccentricity of the gas disc at 540 yr and 2\,700 yr after the passage at pericentre. At the beginning of the simulation the disc is initialized with eccentricity equal to zero. Because of the flyby, the eccentricity of the disc increases significantly shortly after the passage at pericentre. The most spectacular effect is observed for prograde orbits ($\beta0$ and $\beta45$) for which the eccentricity is above $0.3$ beyond $50$ au. Particles with $e>1$ are either accreted by one of the star or become unbound. Polar orbits increase the eccentricity of the outer regions of the disc, but are not able to pump the eccentricity above $0.4$ for the parameters considered here. Retrograde orbits instead are characterized by low eccentricities below $0.1$. The eccentricity is then damped because of the recircularization effects induced by pressure forces \citep{Pfalzner2003}. After 2\,700 yr, we observe that the disc eccentricities remain below $0.2$ within the disc extension.

\subsection{Disc truncation}
\label{sec:truncation}

\begin{table}
\begin{center}
\caption{Final disc size, defined as the radius within which is contained 63.2\% of the disc mass \citep{Bate2018}, for different values of $\beta$, $q$ and $r_{\rm peri}$. These values are to be compared with the initial disc size of 107.4 au and the NoFB case, using the very same definition. $\Delta$~is the difference between the gas disc size $r_{\rm f}$ and the dust disc size $r_{\rm f}^{s={\rm 1 cm}}$. For $\beta$45-q5 this criterion does not work because the disc is heavily stripped. For completeness, we also show the value $r_{\rm f}^{\rm B16}$ obtained using Eq.~\ref{eq:rf}.}
\label{tab:rd}
\begin{tabular}{|c|c|c|c|c|}
\hline
Simulation & $r_{\rm f}$ (au) & $r_{\rm f}^{s={\rm 1 cm}}$ (au) & $\Delta$ (au) & $r_{\rm f}^{\rm B16}$ (au) \\
\hline
        NoFB & 102.0 & 97.8 & 4.2 & x \\
        $\beta$0 & 79.0 & 68.6 & 10.4 & 72.6 \\
	$\beta$45 & 87.2 & 74.0 & 13.2 & 72.6 \\
	$\beta$90 & 100.9 & 92.9 & 8.0 & 72.6 \\
	$\beta$135 & 106.0 & 101.8 & 4.2 & 72.6 \\
	$\beta$180 & 107.9 & 104.2 & 3.7 & 72.6 \\
\hline
         $\beta$45-q02 & 125.0 & 111.3 & 13.7 & 100.2 \\
	$\beta$45-q05 & 105.4 & 91.4 & 14.0 & 84.4 \\
	$\beta$45 & 87.2 & 74.0 & 13.2 & 72.6 \\
	$\beta$45-q2 & 71.2 & 59.4 & 11.8 & 63.2  \\
	$\beta$45-q5 & x & x & x & 52.6 \\
\hline
    $\beta$135-q02 & 122.2 & 116.6 & 5.6 & 100.2 \\
	$\beta$135-q05 & 111.8 & 102.6 & 9.2 & 84.4 \\
	$\beta$135 & 106.0 & 101.8 & 4.2 & 72.6 \\
	$\beta$135-q2 & 102.9 & 99.5 & 3.4 & 63.2 \\
	$\beta$135-q5 & 93.9 & 92.5 & 1.4& 52.6  \\
\hline
    $\beta$45-rp300 & 104.0 & 90.5 & 13.5 & 97.1 \\
    $\beta$135-rp300 & 109.3 & 102.6 & 6.7 & 97.1 \\
\hline
\end{tabular}
\end{center}
\end{table}

As mentioned in Section~\ref{sec:intro}, previous authors have already studied the tidally-induced truncation of discs using N-body codes (shown in the bottom panels of Figs.~\ref{fig:dustr45} and \ref{fig:dustr135}). \citet{Bhandare+2016} proposed the following formula for the tidal truncation radius averaged over inclination angle:
\begin{equation}\label{eq:rf}
  r_{\rm f} = 1.6 \, q^{-0.2} \, \left [r_{\rm peri}/(1\,{\rm au})\right]^{0.72} \,\,\, {\rm au}.
\end{equation}
However, prograde perturbers always produce smaller $r_{\rm f}$ compared to retrograde ones. In other words, the disc final size increases with increasing $\beta$. This can be seen in Figure~\ref{fig:sigmavsr} where we plot the azimuthally averaged gas surface density profile around the primary star 2\,700 yr after the perturber's passage at periastron.

We explore the dependence of the disc truncation on the orbital inclination $\beta$, $q$ and $r_{\rm peri}$, respectively. Here we follow \cite{Bate2018} in defining the characteristic disc radius as the position at which the  enclosed mass reaches 63.2\% of the total disc mass. The main reason for this choice is that we want to estimate the radius that encloses the bulk of the disc. The definition used by \cite{Bhandare+2016} is based on the location of the steepest gradient in the surface density in the outer disc regions. While being closer to the observational estimates (best done by fitting resolved disc images), the complex structures created during the flyby renders this criterion less robust. The disc radii obtained for different values of $\beta$, $q$ and $r_{\rm peri}$ are reported in Table~\ref{tab:rd}, for both the gas and the 1~cm dust discs. Compared to the reference case without flyby (NoFB), the final disc radii for perturbers on retrograde orbits ($\beta$180, $\beta$135) are larger; while for polar orbits ($\beta$90) it is roughly the same. For prograde orbits ($\beta$0, $\beta$45), the final disc size is reduced between $15$ and $20$\%.

We find that the difference between the gas and the dust disc sizes, denoted $\Delta$, increases with increasing $q$ and is larger for prograde orbits compared to retrograde ones. However, caution is required when interpreting the numbers in Table~\ref{tab:rd} since the dust disc size heavily depends on the evolutionary time before the encounter. That is, radial drift would cause more evolved discs to exhibit a more compact dust with higher $\Delta$.

For completeness, we also compute the estimate obtained with Eq.~\ref{eq:rf}. The agreement with $\beta$0 (the most destructive case) is accurate to within $\sim$8\%, but it deviates significantly for higher values of $\beta$. Hence, caution must be used when applying Eq.~\ref{eq:rf} to interpret disc truncation by inclined or retrograde tidal encounters. The disc truncation is relatively unaffected by rotations about the $x$-axis (cf. figure 9 in \citealt{Bhandare+2016}). By varying the value of q for the inclined prograde case (cf. Fig.~\ref{fig:gasdensq}), we retrieve the dependence of $r_{\rm f}$ on $q^{-0.2}$.

\begin{figure}%
    \centering
    \subfloat[Eccentricity 540 yr after the passage at pericentre.]{{\includegraphics[width=0.5\textwidth]{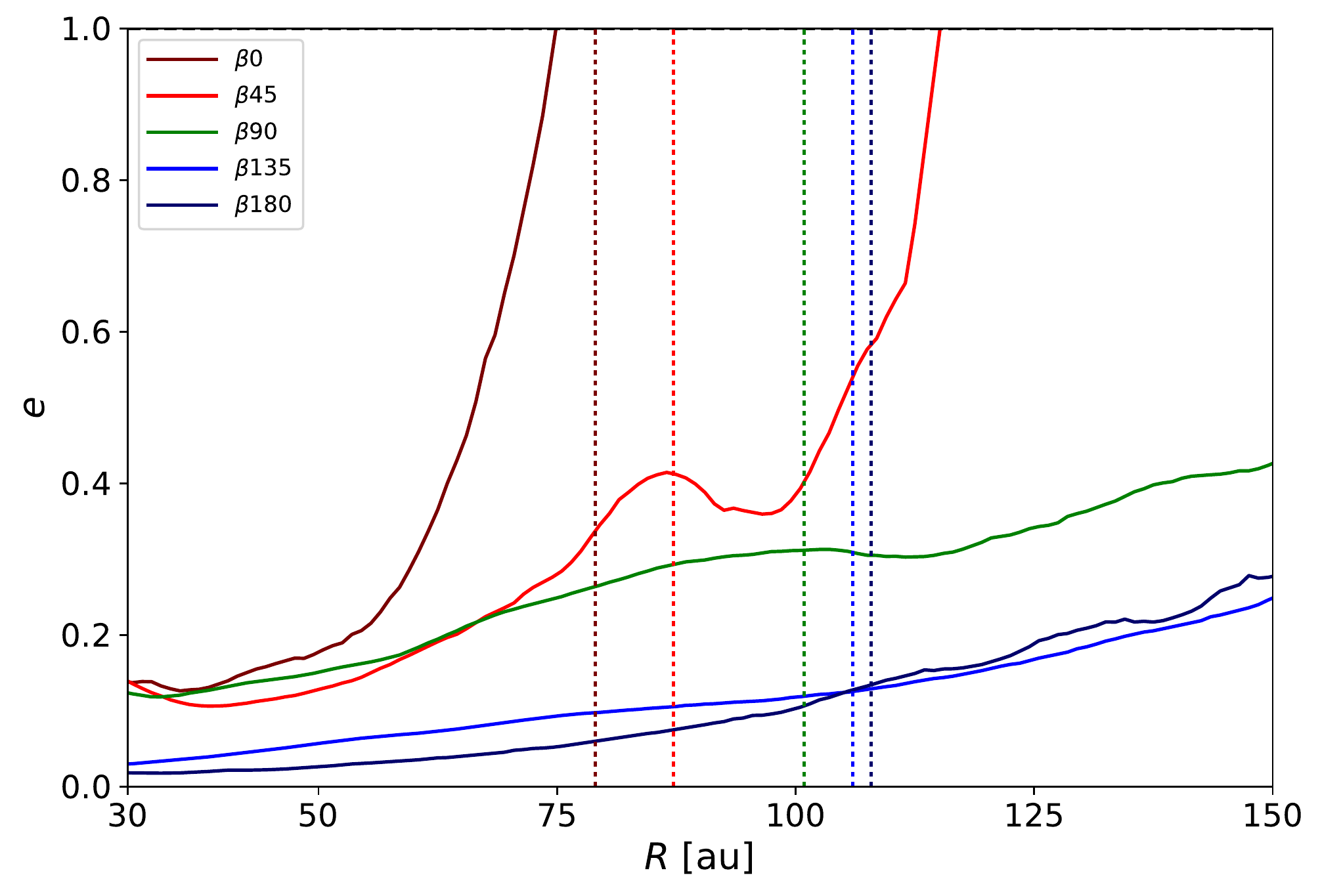} }}%
    \qquad
    \subfloat[Eccentricity 2\,700 yr after the passage at pericentre.]{{\includegraphics[width=0.5\textwidth]{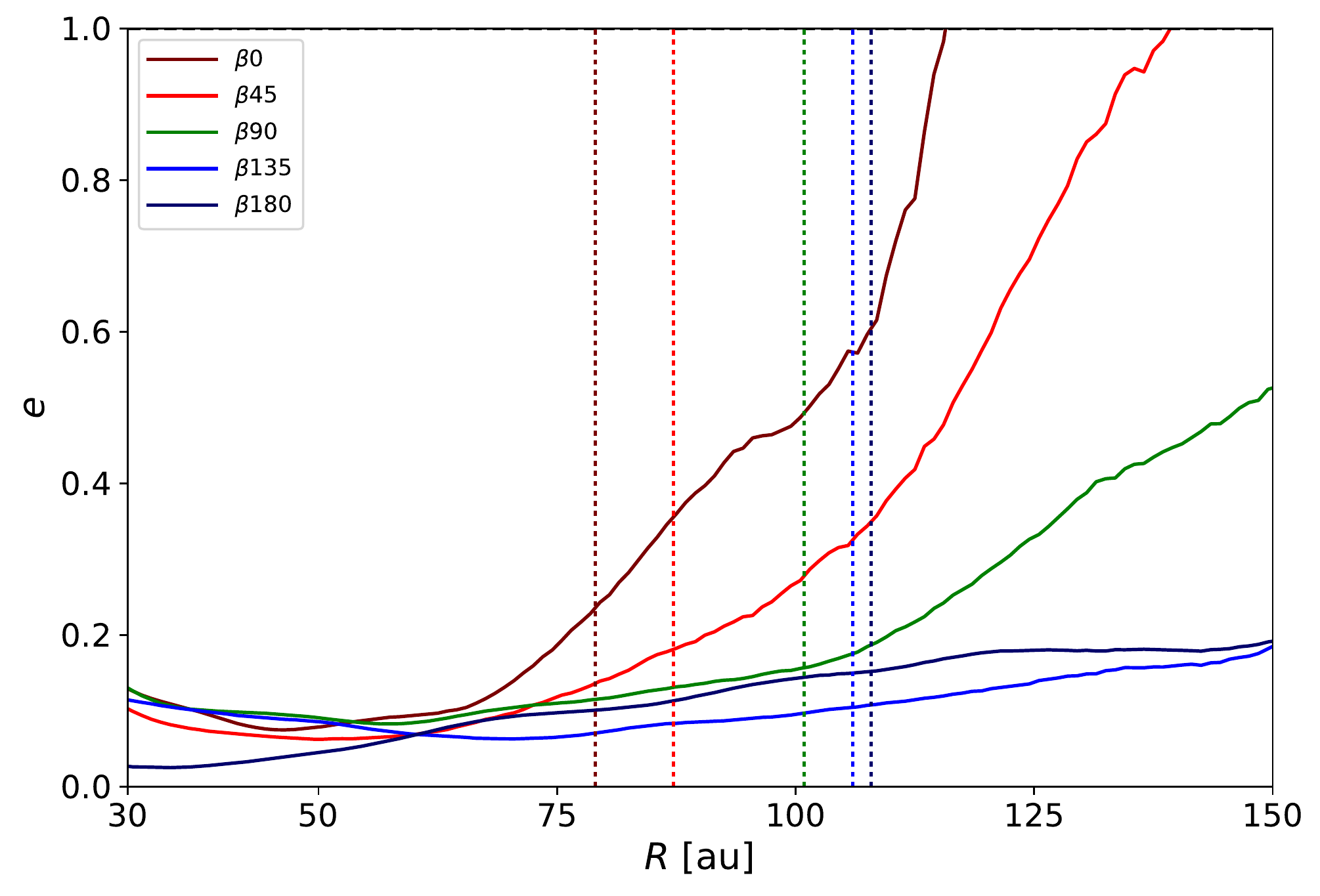} }}%
    \caption{Eccentricity of the gas disc versus radial distance from the central star. Shortly after the passage at pericentre, perturbers on prograde orbits efficiently increase the eccentricity ($e>0.3$) of the outer regions between 50 and 100 au. The particles beyond these radial distances are either captured by one of the star or become unbound. Perturbers on polar orbits lead to eccentricities between $0.2$ and $0.4$. By contrast, retrograde orbits are characterized by low eccentricities ($e\leq0.1$). For later times, the disc eccentricity of the disc remains relatively low within the disc extension ($\sim 30-80$ au).}%
    \label{fig:eccentricity}%
\end{figure}

\begin{figure}%
    \centering
    \subfloat[Azimuthally averaged gas surface density 2\,700 yr after the flyby, for different values of $\beta$.]{{\includegraphics[width=0.5\textwidth]{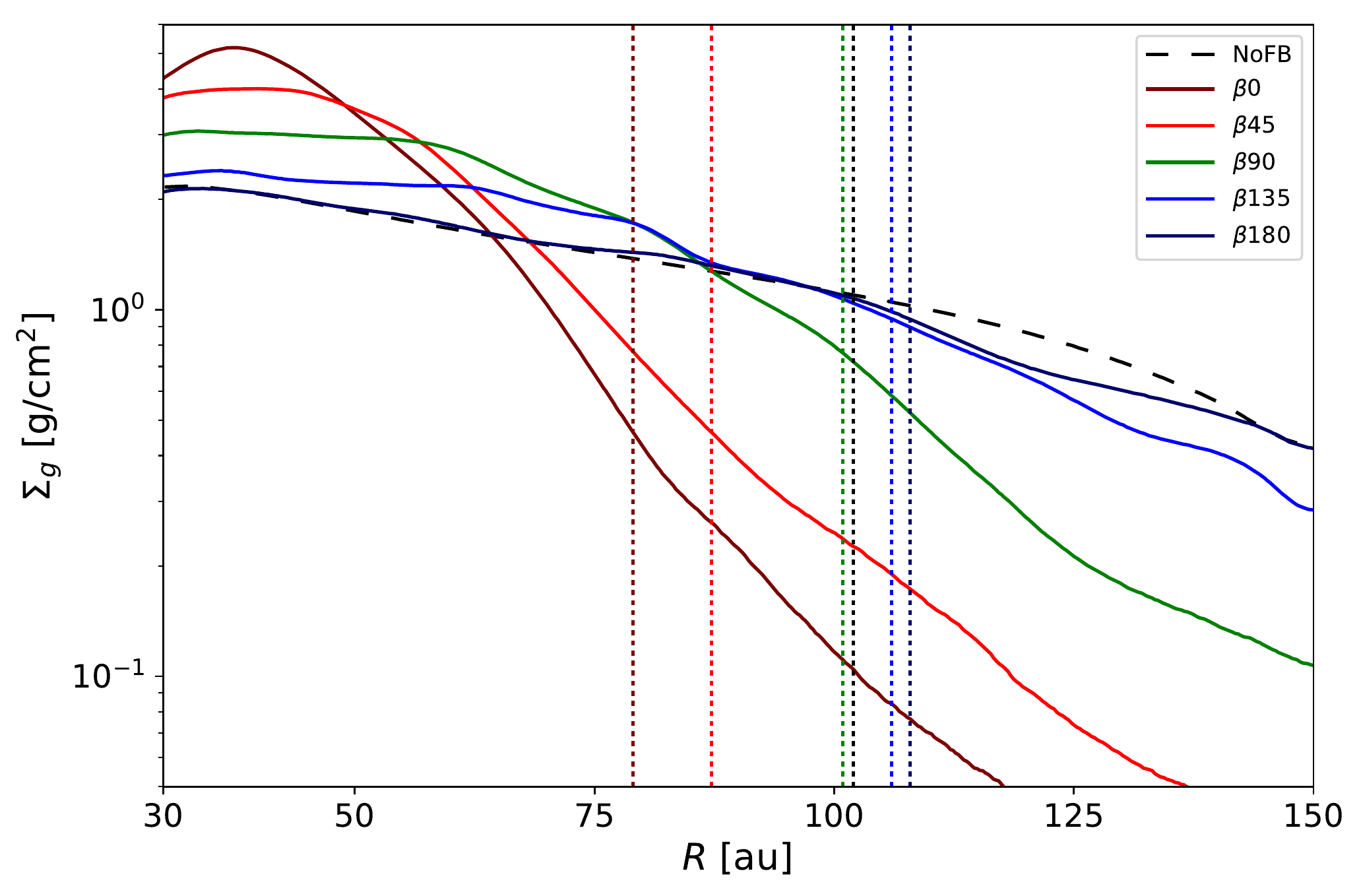} }}%
    \qquad
    \subfloat[Azimuthally averaged dust density profile of 1 cm grains 2\,700 yr after the flyby, for different values of $\beta$.]{{\includegraphics[width=0.5\textwidth]{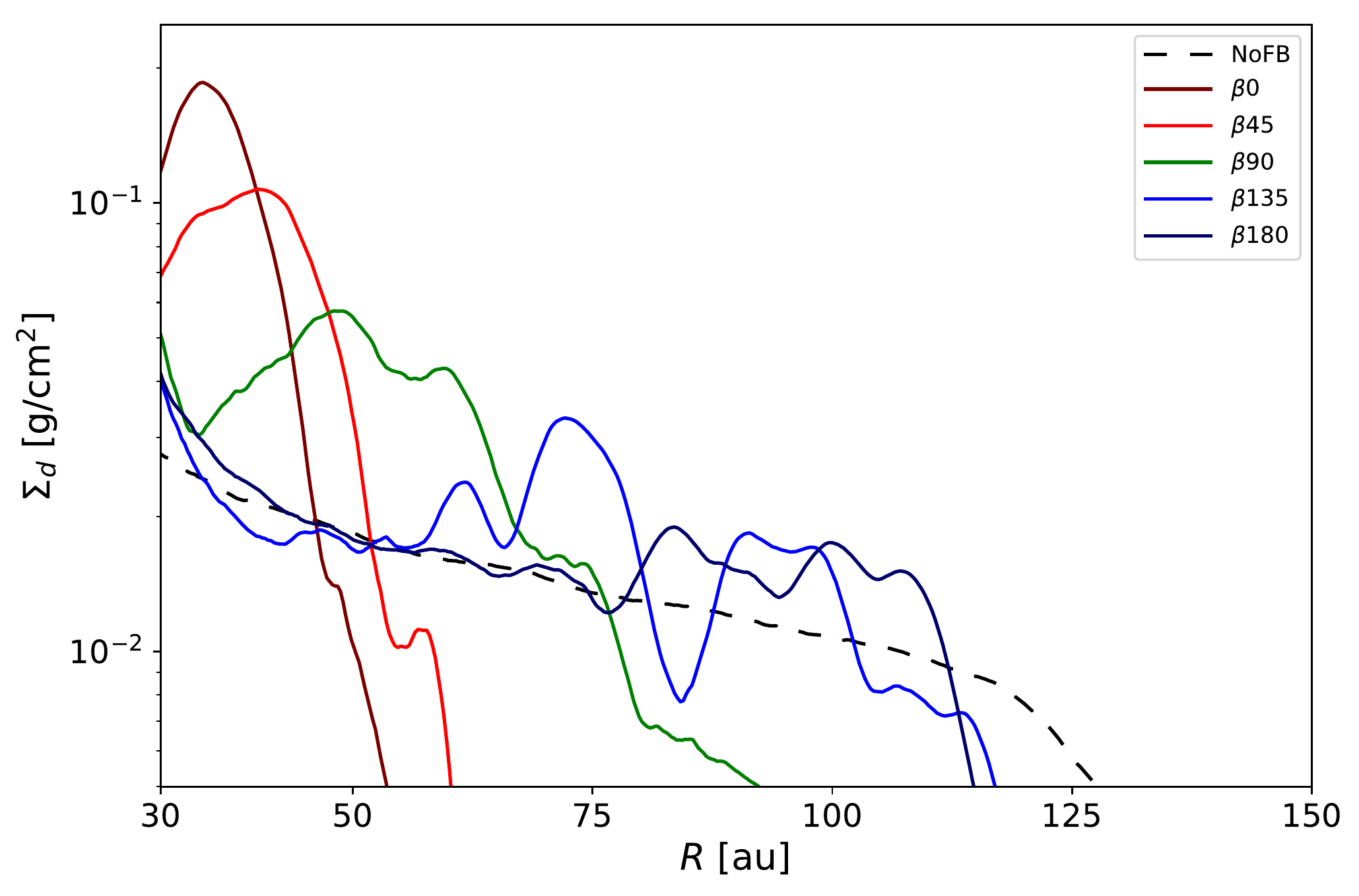} }}%
    \qquad
    \subfloat[Dust-to-gas ratio ($\epsilon$) of 1 cm grains 2\,700 yr after the flyby, for different values of $\beta$.]{{\includegraphics[width=0.5\textwidth]{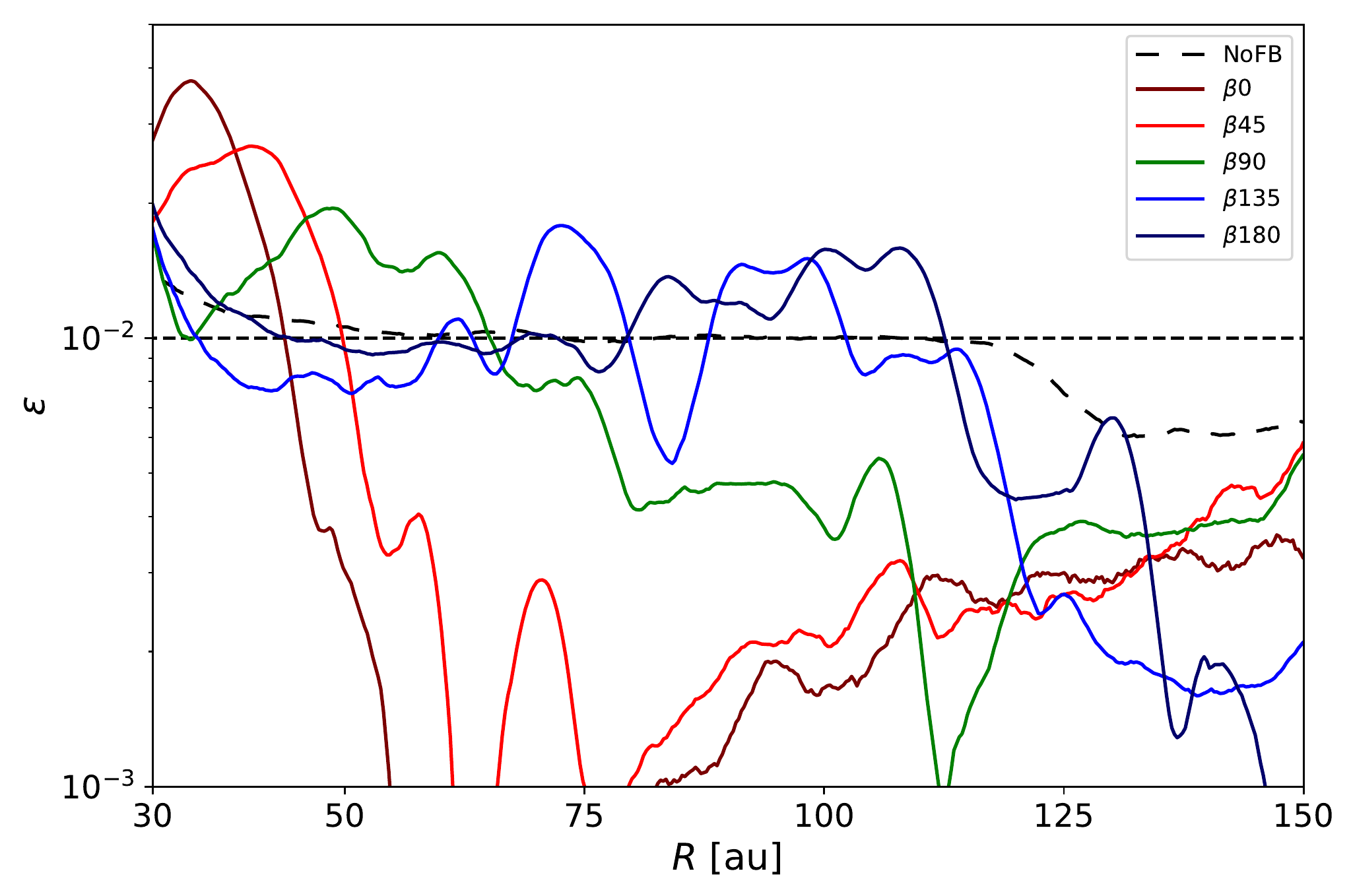} }}%
    \caption{The dotted vertical lines in the top panel correspond to the gas disc sizes obtained with the `63\% disc mass' definition (cf. Table~\ref{tab:rd}). The effect of the flyby is to truncate the disc and to shift the pressure maxima toward the outer regions. These effects are more significant for companions on prograde orbits, compared to retrograde ones.}%
    \label{fig:sigmavsr}%
\end{figure}

Interestingly, by comparing the innermost regions of the disc (below $50$ au) for each configuration, we noticed that the maximum of the gas surface density shifts when the disc is affected by a flyby. We observe this radial shift mainly for prograde orbits. While the regions $\lesssim 30$ au are affected by the inner boundary condition (Appendix~\ref{sec:cavitysink}), for $\beta$0 the peak value of the gas surface density moves from $20$ au up to $40$ au and its value is twice the peak value of the reference case. This occurs because during a prograde encounter the disc material becomes eccentric (cf. Fig.~\ref{fig:eccentricity}) and also loses angular momentum \citep{Ostriker1994, Winter+2018a}. Thus the disc material can either migrate inwards or be captured by the perturber. The moderate to high eccentricities in these regions are responsible for the large inner cavity observed for different values of $\beta$ 2\,700 yr after the encounter. Therefore, in addition to truncation, the radial profile of the gas disc is also affected by the flyby, especially for prograde encounters. Caution is however required when interpreting these results, since the location of the gas pressure maximum depends on the sink size and the resolution as shown in Appendix~\ref{sec:cavitysink}. Increasing the resolution from $10^5$ to $10^6$ particles gives better results in terms of convergence as shown in Figures~\ref{fig:sinktest} and \ref{fig:sinktest-ecc}. Also, the results obtained decreasing the sink radius from $10$ to $1$ au exhibit a similar trend. Hence, we reach a compromise by increasing the resolution while keeping the sink radius to 10 au due to the large computational cost.

Since the dust is coupled to the gas by drag, its dynamics should also be affected by the above effects. Figure~\ref{fig:sigmavsr} shows the 1~cm dust surface density around the primary star 2\,700 yr after the perturber's passage at periastron. The dust disc in every case is more compact than the gas (as expected, Sect.~\ref{sec:dustdensity}). Compared to the reference case, the maxima in $\beta$180 and $\beta$135 are shifted slightly outwards to more than twice the radius of the maximum density in our reference simulation. We checked that this is not a numerical artefact by considering smaller values of the accretion radius ($1$ au instead of $10$ au) for the central star at different resolutions (cf. Appendix~\ref{sec:cavitysink}). 

Figure 11 shows that the dust-to-gas ratio is enhanced by a factor of 2--4 inward of 50 au depending on the orbit. The dust-to-gas ratio increases even further in the regions inward of 20 au but this is mainly an artefact of the inner boundary condition --- in reality this dust would most likely be accreted by the star.

\begin{figure}
\begin{center}
\includegraphics[width=0.5\textwidth]{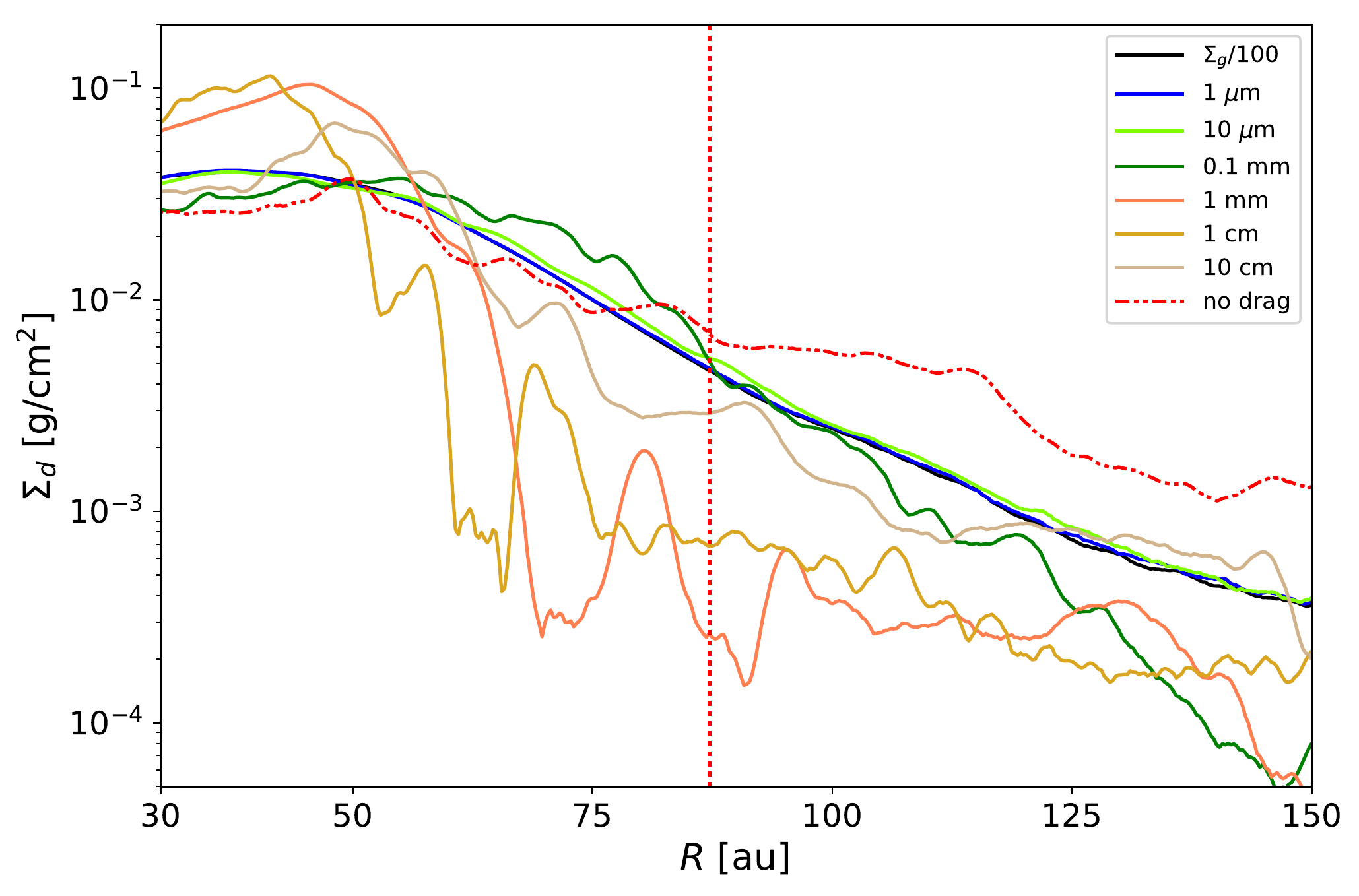}
\caption{Azimuthally averaged surface density of gas (black solid line) and dust with different grain size (colour lines), for $\beta$45, 2\,700 yr after the flyby. The gas surface density is scaled by a factor of 0.01 for direct comparison with the dust phases. Same vertical dotted line as in Fig.~\ref{fig:sigmavsr}. The dust species react differently to the tidal encounter. Besides the tidal truncation, marginally-coupled grains (1 mm and 1 cm) experience an efficient radial-drift.}
\label{fig:sigmacfrrperi}
\end{center}
\end{figure}

Since the radial drift depends on the grain size and density (see Eq.~\ref{eq:stokes}), we expect different disc truncation effects for different dust species. Figure~\ref{fig:sigmacfrrperi} shows the gas and dust surface density, 2\,700~yr after the flyby, for grains of different size in $\beta$45. It can be noticed that the truncation radii vary with grain size. Large grains (i.e. $s$ = [1 mm - 1 cm]) tend to be more concentrated closer to the star due to radial-drift effects. On the contrary, small grains ($s\leq0.1$ mm) are well coupled to the gas. This clearly shows that aerodynamical drag has dramatic effects on dust dynamics during the flyby, which cannot be captured by pure N-body simulations (see Sects.~\ref{sec:dustdensity} and \ref{sec:geomdrag}).

\subsection{Warps}

\label{sec:warps}
\begin{figure}%
    \centering
    \subfloat[Tilt of the disc 2\,700 yr after the flyby for different values of $\beta$.]{{\includegraphics[width=0.5\textwidth]{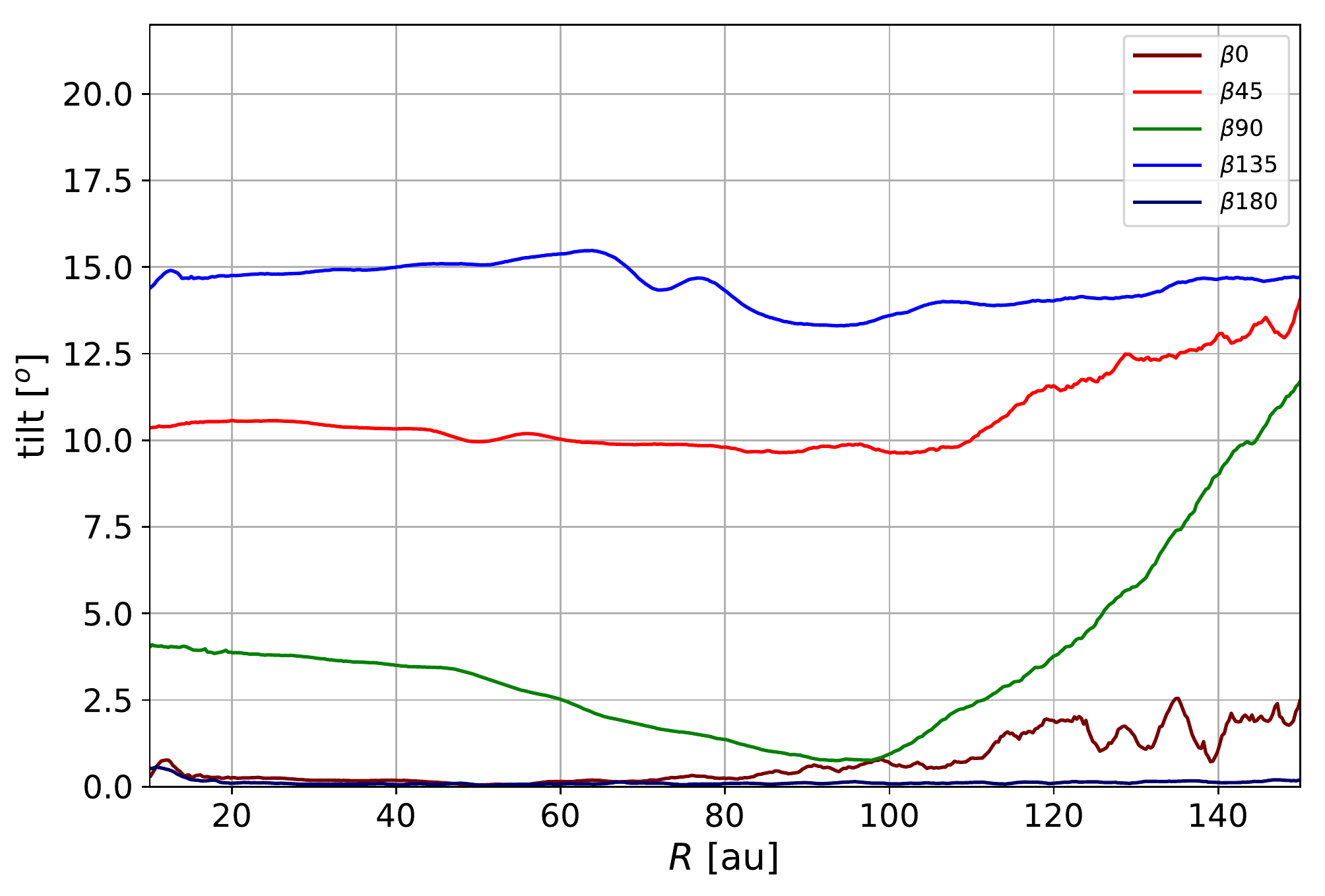} }}%
    \qquad
    \subfloat[Twist of the disc 2\,700 yr after the flyby for different values of $\beta$. For ease of reading and comparison, we plot the opposite of the twist for $\beta$135. Also, a perturber on a polar orbit, but travelling in the opposite direction, produces a twist of opposite sign.]{{\includegraphics[width=0.5\textwidth]{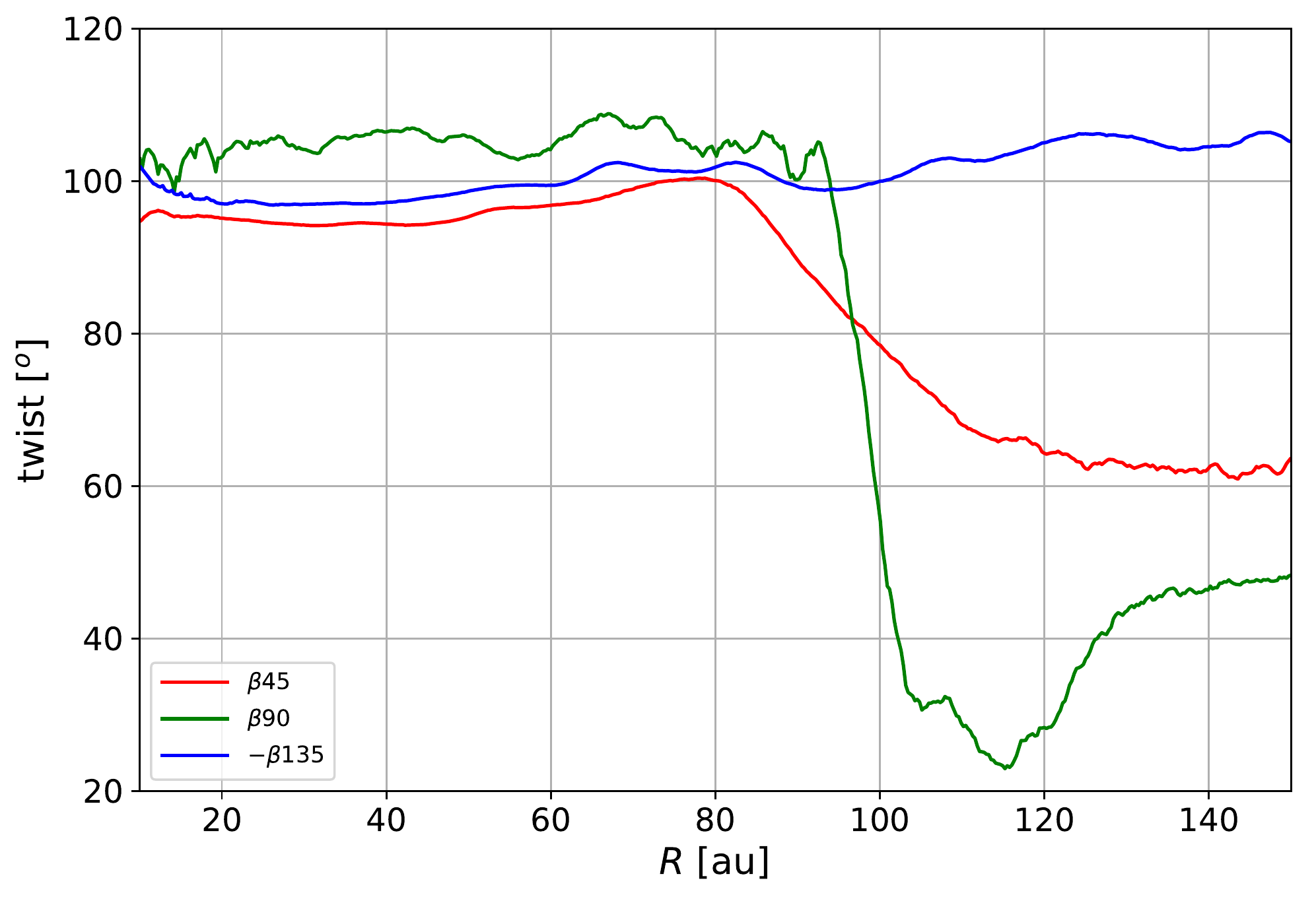} }}%
    \qquad
    \subfloat[Tilt of the disc 2\,700 yr after the flyby for $\beta$135 and different periastron distances: $r_{\rm peri}=100,\,200$ and $300$ au.]{{\includegraphics[width=0.5\textwidth]{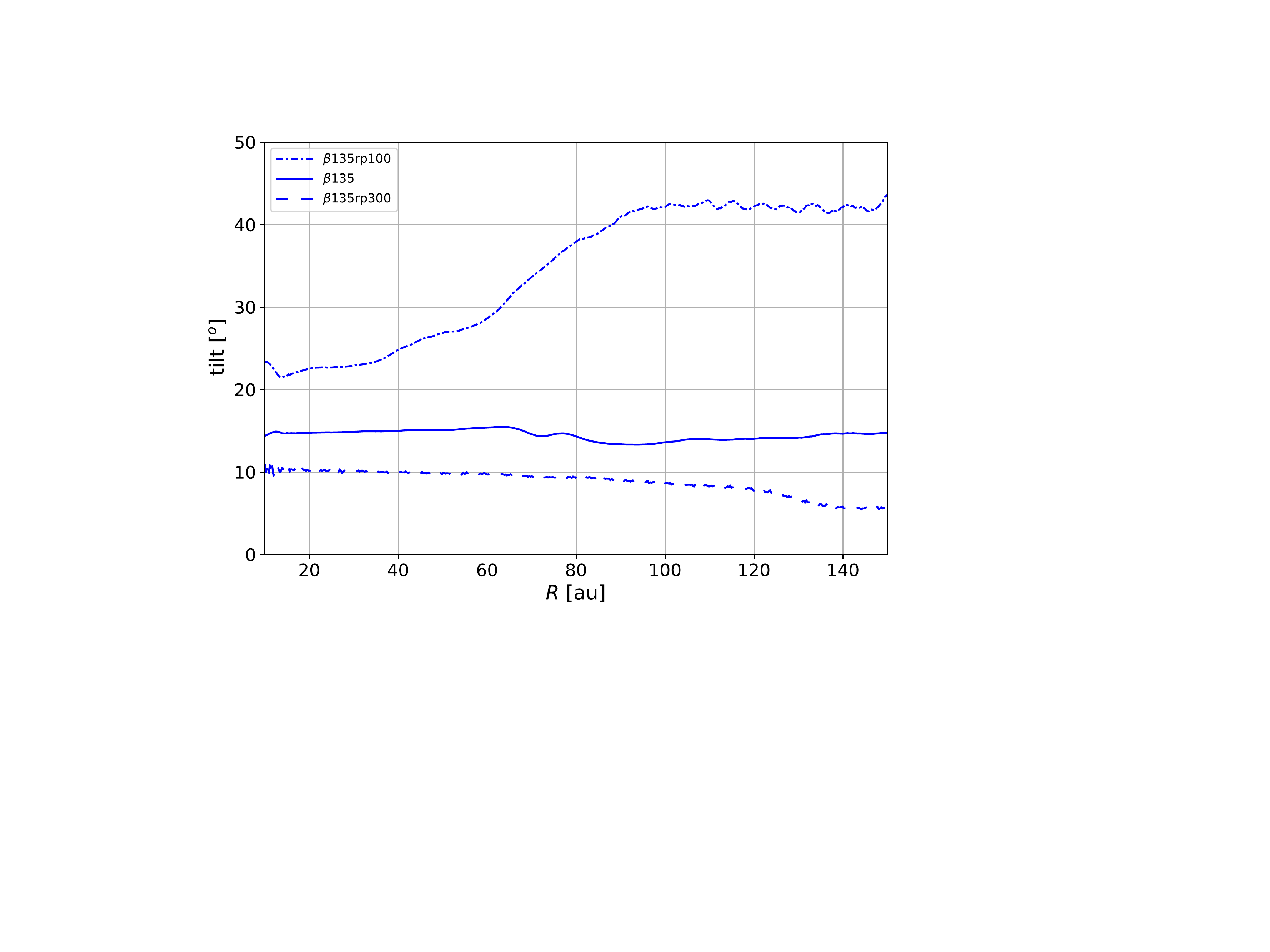} }}%
    \caption{For coplanar encounters, the tilt remains $\sim 0\degree$. A perturber on an inclined retrograde orbit produce the largest tilt ($\sim15\degree$). Coplanar encounters do not modify the twist significantly ($\sim 0\degree$). Perturbers on inclined orbits produce a twist of the disc in the range of $90\degree-110\degree$. Finally, as expected, the tilt increases with decreasing periastron.}%
    \label{fig:tilttwist}%
\end{figure}

\begin{figure*}
\begin{center}
\includegraphics[width=0.9\textwidth]{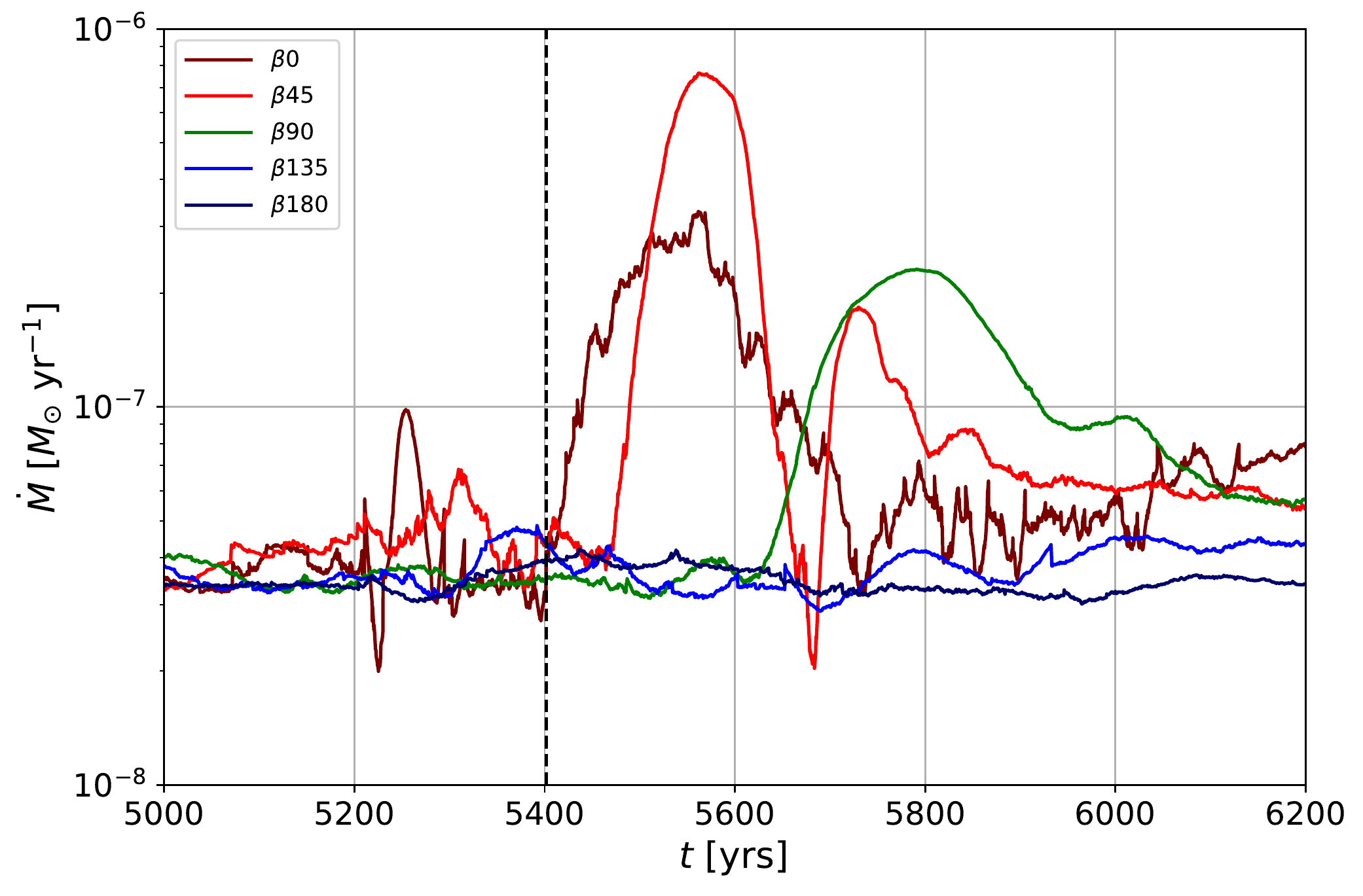}
\caption{Accretion rates onto the primary star during the encounter for different orbital inclinations: $\beta$0 (dark red), $\beta$45 (red), $\beta$90 (green), $\beta$135 (blue), $\beta$180 (dark blue). Shortly after the passage at periastron (vertical dashed line), the primary experiences a dramatic increase in accretion. For prograde orbits, the increase is of one order of magnitude or higher.}
\label{fig:accretion-beta}
\end{center}
\end{figure*}

Our simulations show that during the flyby the gas circumprimary disc can become warped, inclined and twisted. There are two ways of warping a disc: tilt ---  equivalent to the inclination about an axis in the disc plane; and twist --- equivalent to a rotation about the axis orthogonal to the disc plane \citep{ogilvie99a}. If the disc is tilted, then the twist modifies the line of nodes, i.e. the angular position of the ascending node in the reference plane. \cite{XG2016} studied the tilt and twist induced in gas discs by stellar flybys for a broad range of orbital parameters. Here we perform a similar analysis for $q=1$. Figure~\ref{fig:tilttwist} shows the tilt and the twist of the disc 2\,700 yr after the encounter for different values of $\beta$. For the coplanar encounters $\beta$0 and $\beta$180, the tilt remains close to $0\degree$ (as expected). Therefore the twist is not properly defined for these configurations. For $\beta$0 the tilt is slightly above $0\degree$ beyond $100$ au, but the density in these regions is very low (cf. Fig.~\ref{fig:sigmavsr}). The bulk of the disc is within the truncation radius reported in Table~\ref{tab:rd}.

The inclined cases are more interesting: As in \cite{XG2016}, we observe that a perturber on an inclined retrograde orbit is the most efficient at tilting the disc. For the orbital and disc parameters considered here ($r_{\rm peri}=200$ au, $r_{\rm out}=150$ au, $q=1$, $e=1$), the disc inclination is equal to $10\degree$ and $15\degree$ for $\beta$45 and $\beta$135, respectively. The polar case $\beta$90, shows a moderate inclination for radii below $100$ au, and an increasing inclination for the material in the outer regions. However, the regions beyond $100$ au are beyond the truncation radius. Therefore, our simulations show that, while prograde encounters are more efficient at truncating and capturing disc material, retrograde encounters are the most efficient at tilting the disc.

Regarding the twist, inclined prograde ($\beta$45), polar ($\beta$90) and inclined retrograde ($\beta$135) orbits produce twists equal to $+90\degree$, $+90\degree$ and $-90\degree$, respectively. If the perturber moves along the polar orbit the other way around (i.e. $\beta$270), the twist is then equal to $-90\degree$. By rotating the orbit about the $x$-axis, i.e. for orbits where the pericentre location is out of the disc plane (not shown here), intermediate values for the twist can be obtained. Therefore, a perturber on an inclined orbit is able to efficiently tilt and twist the disc. In Appendix~\ref{sec:appendix-kin-q}, we show the warping increases with increasing $q$ (cf. Figs.~\ref{fig:tiltq} and ~\ref{fig:twistq}).

Last but not least, if we change the periastron distance for $\beta$135 (i.e. the case with the highest tilt), then this also modifies the tilt (bottom panel in Fig.~\ref{fig:tilttwist}). If $r_{\rm peri}=100$ au, then the tilt is roughly $23\degree$ for the inner regions; whereas if $r_{\rm peri}=300$ au, then the tilt is equal to $10\degree$. The twist in this case shows less variability and ranges from $90\degree$ to $105\degree$. So the perturber, even at $150$~au from the disc outer edge, can significantly warp the disc around the primary, provided that the perturber is on an inclined retrograde orbit.

\subsection{Mass accretion and outbursts}
\label{sec:accretion}

Figure~\ref{fig:accretion-beta} shows the accretion rate onto the primary star as a function of time for various values of $\beta$, assuming $q=1$ and $r_{\rm peri}=200$ au as before. Several hundreds of years before periastron passage, the accretion rate $\dot M$ oscillates between $3$ and $5\times 10^{-8} \, M_{\odot}\,{\rm yr}^{-1}$. Just after periastron, $\dot M$ increases significantly for prograde orbits ($\beta$0 and $\beta$45) and undergoes large fluctuations over a time-scale of several hundreds of years. The highest accretion rate is reached for the inclined prograde orbit: a rate $\sim 20$ times higher than before the perturber's passage at periastron. For coplanar prograde and polar orbits, the increase is only of a factor $10$. For a polar perturber there is a delay in accretion of roughly 300 yr compared to prograde ones. By contrast, for retrograde perturbers ($\beta$135 and $\beta$180) the accretion rate remains almost unchanged showing little variability during the encounter. We found this to remain true for non-penetrating encounters, i.e. $r_{\rm peri} > R_{\rm out}$.

The efficiency of the eccentricity pumping of the outer disc regions for different values of $\beta$ is responsible for this trend. Figure~\ref{fig:eccentricity} shows that for $\beta0$ and $\beta45$ a large fraction of the disc ($r>50$ au) reaches high eccentricities ($e>0.4$) shortly after the encounter. This means that, after the pericentre, a significant number of particles get close enough to the central star to be accreted, which leads to the accretion event. On the contrary, for retrograde orbits, the disc eccentricity remains relatively low ($e<0.2$), which explains the low accretion rates observed in Fig.~\ref{fig:accretion-beta}. Again, caution is required when interpreting these results, since the accretion reported strongly depends on the amount of matter available close to the star, the central star sink radius and the resolution (Appendix~\ref{sec:cavitysink}). Numerically, the stellar accretion rate corresponds to the sink accretion rate, which varies with the sink size. Given the vast space of parameters of this study, it is prohibitively expensive to present fully converged results for the accretion rate. Either way, flybys are expected to trigger outburst events regardless of the specific numerical setup. In Appendix~\ref{sec:cavitysink}, we show the accretion rates for $\beta45$ --- where the strongest accretion is observed --- for different sink radii (1 and 10 au) and resolution ($10^5$ and $10^6$).

Additionally, we observe that with larger $q$ and/or smaller $r_{\rm peri}$, $\dot{M}$ is higher (see Fig.~\ref{fig:accretion-q}). Also, provided that the accretion rate is proportional to the disc mass \citep{Clarke&pringle2006}, then by increasing the mass by a factor 100, the accretion rate is increased by the same factor. Thus, considering heavier or lighter discs the accretion rate can be increased or reduced, respectively. As discussed in Sect.~\ref{sec:flybymechanism}, these sudden accretion events can be directly linked to FU Orionis stars, in which a spectacular and highly variable increase in stellar luminosity is observed over a short period of time \citep{Vittone&Errico2005}.

\section{Discussion}
\label{sec:discussion}

\subsection{Flybys as a symmetry-breaking mechanism}
\label{sec:flybymechanism}

We showed that a variety of disc structures --- spirals, bridges and warps --- appear during a stellar flyby, both in the gas and the dust distributions. The results obtained for the gas phase agree with previous works (e.g. \citealt{Clarke&Pringle1993}, \citealt{Dai+2015}, \citealt{XG2016}). The novelty of our approach is that we also studied how the dust responds to this kind of tidal encounter. The aerodynamic coupling between dust and gas means that the response of the dust disc should be a strong function of grain size (cf. Sect.~\ref{sec:dustdensity}). Dust grains with a Stokes number close to $1$ ($s \approx 1$ mm for the disc model considered here) experience the most efficient radial drift. This translates into a more compact dust disc prior to the encounter, for which the flyby parameters are different from those of the gas disc. Hence, when the disc is tidally perturbed, dust and gas do not respond in the same way. Comparing the gas and 1~cm dust distributions of Figs.~\ref{fig:gas-panel} and \ref{fig:dust-panel}, we observe that the spirals in the dust are sharper (i.e. narrower) and less radially extended than the ones in the gas. This is due to the combined effect of the gas drag and the gravitational perturbation during the tidal encounter.

 Our models have important implications for disc observations which trace different dust populations at different wavelengths. For example, Near-Infrared (NIR) or scattered light observations are sensitive to the distribution of micrometric grains, while dust thermal continuum observations at mm-radio wavelengths trace larger dust grains (100 $\mu$m < $s$ < 1 cm). These observations provide information on the surface layer and the mid-plane of the disc, respectively.
 
 RW Aur is an example of a system where both kinds of observations show strong differences \citep{Cabrit+2006,Rodriguez+2018}. Both RW Aur A and B show evidence of circumstellar discs. The two discs are compact (15-20 au) and misaligned to each other. There is also a prominent spiral arm around RW Aur A, which is likely due to a tidal encounter as suggested by \cite{Dai+2015}. Considering this scenario, the authors fit many of the flyby parameters of RW Aur through hydrodynamic models and synthetic observations. Specifically, the main morphological and kinematic features of RW Aur were well reproduced with a single parabolic ($e=1$) flyby. However, \cite{Rodriguez+2018} suggested that the morphology of RW Aur is more likely caused by multiple eccentric flybys, because of the multiple tidal streams observed. Additionally, the probability of a recent flyby as the one suggested in RW Aur is low given the low stellar density in Taurus. This is the main argument in favour of a bound companion.

 HD135344B --- separated by 21 arcsec (in projection) from HD135344A --- is also a remarkable system: scattered light observations revealed spiral structures in the outer disc \citep{muto12a,garufi13a,stolker16a}, while \mbox{(sub-)}mm ALMA observations showed a large-scale horseshoe in the dust continuum \citep{perez14a,pinilla15a,van-der-marel16b}. The effects of a (yet undetected) planet at 30 au in the disc could explain the puzzling disc morphology. However, a (retrograde) stellar flyby and the resulting differential tidal response of both phases could explain some of these features (compare the top and middle panel of Fig.~\ref{fig:dustr135}). However, in HD135344B, the flyby is unlikely to have caused the inner disc cavity in the dust. More generally, flybys should be considered viable explanations for discs with spirals and different dust and gas distributions, but where no nearby companions have been detected.

Lastly, Figs.~\ref{fig:accretion-beta} and \ref{fig:accretion-q} show that the accretion rate of the primary can suddenly increase by a factor of 10 or more during a stellar flyby. The resulting outburst --- if interpreted as a FU Orionis event --- would last for time-scales of the order of a hundred years. Therefore, assuming a tidal encounter occurs, we can link the stellar brightness increase (5-6 mag) with the companion's gravitational perturbation. This idea was originally proposed by \cite{BonnellBastien1992} and has been then further investigated from both observational \citep{Aspin&Reipurth2003,Reipurth&Aspin2004,Greene+2008} and  theoretical \citep{Pfalzner2008} perspectives. Remarkably, FU Ori itself --- the prototype of FU Orionis objects --- is a system composed of two stars \citep{Wang+2004}: FU Ori of $0.3 \, M_\odot$ and FU Ori S of $1.2 	\, M_\odot$ \citep{Beck&Aspin2012}. Also, both stars have protoplanetary discs, which are likely interacting with each other as suggested by the recent observations by \cite{Hales+2015} and Principe et al. (in prep.). The fraction of FU Ori objects with candidate companions remains however hard to estimate --- observations at higher spatial resolution are needed. Nevertheless, flybys can efficiently trigger fast accretion events. Our results indicate that higher accretion rates are obtained for \textit{inclined} prograde orbits and high mass ratios.

The flyby scenario is similar to the `encounters of type 1 and~2' in triple systems proposed by \cite{Reipurth+2014}, where accretion events occur at each periastron passage. In the former, the major brightening is due to the tidal perturbations experienced by the circumstellar material around each star; while the latter, occurs when a binary has formed and the third star orbits around it, triggering episodic accretion events. 

\subsection{Tidally-induced spirals}
\label{sec:spirals}

Recently, a large number of spirals have been reported in discs around T~Tauri and Herbig AeBe stars. These features are usually detected in scattered light at near-infrared wavelengths \citep{muto12a,grady13a,garufi13a,wagner15a,reggiani17a}. There are also three cases where spirals were detected by other means: in HD142527 through molecular CO line emission \citep{christiaens14a} and in Elias 2-27 and L1448 IRS3B in the dust continuum sub-mm emission \citep{perez16a,tobin16a}, all three with ALMA.

Although limited by the resolution of the observations, two almost symmetrical spiral arms seems to be most common pattern in all the discs where a spiral structure is detected. These arms are frequently open, with pitch angles of the order of 10\degr ~to 20\degr. Interestingly, some objects exhibit different substructures at different wavelengths, making their interpretation difficult (e.g. \citealt{boehler18a}). Several mechanisms have been proposed to create spirals in protoplanetary discs: gravitational instabilities \citep{kratter16a}; planetary/stellar companions embedded in the disc \citep{dong15a,Price+2018}; external massive perturbers \citep{Pfalzner2003,quillen05a,Dai+2015}; accretion from an external envelope \citep{Harsono+2011,Lesur+2015,Hennebelle+2016,Hennebelle+2017}; and finally, asymmetric stellar illumination patterns \citep{montesinos16a,montesinos18a}. Flyby-triggered spirals are transient (lasting a few thousand years), whereas spirals triggered by accretion and gravitational instability are long-lived. Also, spirals caused by gravitational instabilities have constant pitch angles and, for a given disc, all the spiral arms can be fit by the very same equation. By contrast, spirals caused by external perturbers have pitch angles that increase with increasing distance to the central star. Each of the two spiral arms in this case should be fit individually \citep{Forgan+2018}.

Given the age and the low mass of the discs where spirals have been detected, these are unlikely to be produced by self-gravity \citep{lodato04a,dipierro14a}. Therefore, the most widely accepted interpretation is the presence of a planetary or low-mass stellar companion orbiting external to the spirals in the disc \citep{dong15a}. However, in order to match the observations \citep{muto12a,benisty15a}, strong assumptions need to be made. For instance, using the linear density wave theory, very high disc temperatures --- several hundreds of K --- are required at the companion's location to reproduce the pitch angle of the spirals \citep{rafikov02a,zhu15a}. The vertical temperature gradient of the disc also affects the pitch angle of the spirals: these are more open in the hotter (upper) layers compared to those in the (cooler) mid-plane of the disc \citep{Juhasz&Rosotti2018}. However, this is unable to alleviate the tension between theoretical models and observations. 

Another issue of this scenario is the difficulty in forming of giant planets at large stellocentric distances (several tens or hundreds of au), which is at odds with the core accretion theory. Mitigating this, recent ALMA discoveries show that such planets indeed exist \citep{Pinte+2018}.

Flybys provide a means to induce long-lasting, prominent spirals in relatively low-mass discs. Our results show that these spirals form both in the gas and in the dust during the encounter (especially for prograde orbits) and that they are able to survive for a few thousands of years. Furthermore, by inclining the orbit with respect to the disc plane, it is possible to generate spirals that appear to have larger pitch angles compared to coplanar encounters. 

\cite{Pfalzner2003} showed that there are two mechanisms to form spirals during coplanar prograde encounters: one due to the gravitational perturbation and another caused by the movement of the primary star with respect to the centre of mass. The former induces a shock-like pattern characterized by a jump in velocity around the angular position of the spiral (from sub-Keplerian to super-Keplerian). The latter, by contrast, does not produce such a jump. This is also observed in our simulations, where prograde and polar encounters produce two spirals with different pitch angles, in agreement with \cite{Pfalzner2003}. However, retrograde encounters, for which the perturbation time is shorter, tend to produce a spiral of the second kind only. This is better seen for the dust in the $\beta$135 simulation in Fig.~\ref{fig:dust-panel}, where the most prominent spiral is on the left in our figure. Interestingly, none of the spiral arms in this particular case are pointing towards the perturber as is the case in \cite{dong15a}, for an external but bound companion/planet.

 Despite the broad range of spirals for different flyby parameters, our dynamical catalogues presented in Figs.~\ref{fig:gas-panel}-\ref{fig:kinematics} can be used to infer the orbit of the perturber. In the case where the perturber is not yet detected, following the spirals could indicate where to look for it. On the other hand, if it was detected, the morphology of the spirals can constrain its orbital parameters. Based on our results, this methodology has already been successfully applied to observations of highly-asymmetrical systems such as UX Tau (M\'enard et al., in prep.) and FU Ori (Principe et al., in prep.). 

\subsection{Disc truncation}
\label{tidaltruncation}

A second `smoking gun' for flybys is the tidal truncation of the disc. As shown in Sect.~\ref{sec:truncation}, prograde encounters are more efficient at truncating and capturing disc material. Using Eq.~\ref{eq:rf}, for a given $M_1$ and a given $r_{\rm peri}$, then $r_{\rm f} \propto M_2^{-0.2}$. Hence, as expected, by increasing the $M_2$ or decreasing $r_{\rm peri}$ it is possible to truncate the disc more severely (cf. Table~\ref{tab:rd} and Appendix~\ref{sec:appendix-kin-q}). Nevertheless, the orbital inclination of the perturber also modifies the truncation radius by several tens of au (Sect.~\ref{sec:truncation}). Therefore, Eq.~\ref{eq:rf} should not be applied directly when interpreting observations if the inclination is unknown.

The truncation of the dust disc induced by the flyby should be distinguished from the reduction of disc size due to radial-drift only \citep{W77}. After a given evolutionary time, the dust disc will become more compact than the gas one: $r_{\rm dust} < r_{\rm gas}$. This effect depends on the grain size. If we assume that the dust particles have efficiently drifted prior to the encounter, then the dust disc will not be truncated in the same way as the gas disc during the flyby. This effect can be seen in Fig.~\ref{fig:sigmavsr}. Therefore, more evolved discs can lead to more radially-compact dust structures after the flyby. Moreover, if a planet is already present in the disc, it may efficiently stop the inward drift of the dust. In such a case, the resulting planetary gaps should in principle be detectable in the mm continuum with ALMA \citep{Gonzalez+2015}.

 For a massive enough companion (or equivalently a penetrating enough periastron), it is possible to truncate gas and dust discs at the same radius. If this occurs, the radial drift is effectively reset but with increased dust-to-gas ratio. Almost all the truncated material will then be in the form of gas and small particles, whereas most of the large grains will remain bound to the primary star. By varying $r_{\rm peri}$ from penetrating to far encounters, \cite{Bhandare+2016} found that the disc is truncated for periastron distances as large as $5$ times the initial disc size. Therefore, the discrepancy in size between gas and dust discs could be ascribed to (relatively distant) stellar flybys in the past. According to when the encounter happens, the disc sizes in both phases will be different because of previous disc evolution (viscosity and radial-drift). For instance, the longer the evolution before the encounter, the more compact the dust disc at the time of periastron. However, viscous evolution alone can naturally produce this feature. In this regard, flybys are not a necessary condition to account for differences in disc sizes in the NIR and in the continuum.
 
External photoevaporation can also efficiently truncate the disc on timescales of order several Myr, as shown by \cite{Winter+2018b}. This occurs when the disc is embedded in an environment where the UV flux from massive stars is strong, i.e. in regions of high stellar density ($>10^4$ stars pc$^{-3}$). This is precisely the environment where flybys are expected to be the most likely, also leading to disc truncation. However, \cite{Winter+2018b} find that photoevaporation is always the dominating effect for disc truncation in these environments. Nevertheless, the dynamical response of the disc is radically different between both scenarios. In particular, none of the disc asymmetries shown in Sect.~\ref{sec:results} are expected in the case of discs experiencing photoevaporation without encounters.

\subsection{Mass transfer and dynamical fingerprints}
\label{sec:masstransfer}

Stellar encounters can enable transfer of material from one star to the other, especially for prograde encounters. On this basis, we suggest the existence of a class of young stars with dynamical ``anomalies'' due to exchange between two discs. For simplicity, in this work we neglected the presence of a pre-existing circumsecondary disc. However, \cite{Pfalzner+2005} showed that during disc-disc encounters the exchange of material is mainly \textit{additive} (in N-body calculations at least). This means that the material transferred from the primary disc to the perturber is not affected by the presence of a potential secondary disc (and vice versa). These dynamical signatures in discs are likely below detection thresholds, but are an expected dynamical outcome nonetheless. In the context of debris discs, this aspect has been recently addressed by \cite{Jilkova+2016}. Interestingly, this study estimates the distribution of the orbital parameters for the captured material as a function of the flyby parameters.

\cite{Pfalzner+2005} showed that material is captured mainly for prograde encounters, close pericentre distances and massive perturbers. This is in agreement with our results (cf. Figs.~\ref{fig:gas-panel}-\ref{fig:dust-panelz} and \ref{fig:r45q-dens}). For instance, we observe the highest capture of disc material by the perturber for $\beta45$ and it reaches approximately 15\% of the initial disc mass. Interestingly, the capture of material and whether the disc is heavily stripped or not provides useful information about the inclination of the perturber's orbit. For instance, if we suspect a flyby in a given system, then by looking at the disc structure it is possible to distinguish between prograde and retrograde configurations. This approach was already followed by \cite{Dai+2015} for RW Aur.

Finally, perturbers on hyperbolic orbits are less efficient in capturing material compared to parabolic ones as shown by \cite{Larwood&Kalas2001} and \cite{Pfalzner+2005b}. Both works study the mass transfer between a debris disc (without viscous and pressure forces) and a stellar perturber. Therefore, they constitute an upper limit for the mass transfer expected between a protoplanetary disc and a stellar perturber.

\subsection{Dust: accelerated drift, mixing and collisions}
\label{sec:drift}

In Sect.~\ref{sec:truncation}, we showed that after the flyby the surface density maximum in the gas distribution is shifted outwards, especially for prograde encounters (cf. Fig.~\ref{fig:sigmavsr}). In addition, the gradient of the density profile becomes steeper. This has direct consequences for the radial-drift of the dust particles, as shown in detail by \cite{Laibe2012}. Dust grains drift towards the pressure maxima. By comparing the reference case (NoFB) to $\beta$0 for instance, we notice that the radial position of the maximum density is shifted. For retrograde encounters ($\beta$135 and $\beta$180), the peak is less shifted but the peak values are higher compared to the prograde cases. Hence, besides the truncation of the disc, the flyby accelerates the radial-drift of the dust towards the (shifted) pressure maxima in the disc. This effect might potentially trigger streaming instability in the disc \citep{Johansen&Youdin2007} and/or self-induced dust traps \citep{Gonzalez+2017} by quickly producing regions of increased dust-to-gas ratio (cf. bottom panel in Fig.~\ref{fig:sigmavsr}).

In addition to this radial concentration, dust can also be trapped inside spiral arms. This is seen for example in Figs.~\ref{fig:dustr45} and \ref{fig:dustr135}, where the spirals are more or less sharp according to grain size. This means that spirals behave as dust traps, but that they also mix grains of different sizes and composition \citep{Pignatale+2017}. This might favour their growth since collisions among unequal-sized grains tend to be more constructive than for equal sizes \citep{Blum2018}. This is not expected to occur in the absence of a flyby. Additionally, given that the flyby perturbation can increase the eccentricity and/or the inclination of the particles in the disc, this can lead to an increase in the rate of collisions.

Hence, stellar flybys may strongly affect planet formation in protoplanetary discs around young stars. In Sect.~\ref{sec:dustdensity}, we showed that flybys can create regions of increased dust-to-gas ratio ($\epsilon~\sim~0.1$). Given that dust experiences efficient vertical settling, $\epsilon$ is expected to be even higher at the mid-plane: $\epsilon(z=0)>\epsilon$. In isolation, this naturally occurs close to the star at a few au. However, due to the flyby, $\epsilon$ is dramatically increased at several tens of au ($\sim$20-40~au). Even if viscosity may re-circularize and smooth the perturbations in the gas on a few orbital time-scales, this does not happen as quickly for the tidally-induced regions of high $\epsilon$ in the disc. Even better, viscosity may dilute the density maxima of the gas, locally increasing $\epsilon$. At such high values of $\epsilon$, the back-reaction slows down the radial-drift and the viscous drag of the particles significantly. A stellar flyby may thus generate favourable conditions for planet formation far away from the star. As encounters most likely occur during the early evolution of molecular clouds, protoplanetary discs should then bear flyby signatures. This suggests that the planetary architecture around a significant fraction of stars may be connected to the stellar cluster environment. If true, given that flybys are highly stochastic, we should expect a large diversity of planetesimals and planets as flyby by-products.

\section{Conclusions}
\label{sec:conclusions}
We have explored the dynamical signatures of protoplanetary discs perturbed by stellar flybys. The tidal interaction with the perturber produces spirals, bridges, warps and eccentric discs. Given that such stellar flybys are more likely to occur on inclined orbits, this naturally leads to non-coplanar disc geometries. Comparing our results to recent disc observations seems promising. In particular, we propose flybys as an alternative scenario for discs with spirals where, despite an active search, no external/internal companion has been detected. As discussed, flybys are not the only explanation --- being infrequent --- but may be a viable alternative in some cases. The tidally-induced patterns survive for thousands of years, easily allowing the perturber to leave the field of observation. Hence, we may be observing disc structure caused by historical flybys of stellar companions.

Our main results can be summarized as follows:
\begin{enumerate}
\item Prograde orbits of the perturber are more destructive than polar and retrograde orbits. This translates into more heavily truncated discs and more material captured by the perturber for prograde configurations. However, less prominent dust and gas spirals can still form for polar and inclined retrograde orbits.
\item Tidally-induced spirals show large pitch angles during the encounter ($20$--$30\degree$), which subsequently moderate to $\sim 10$--$20\degree$). These spirals occur out of the disc plane if the perturber is on an inclined orbit. Increasing the mass ratio $q$ or decreasing the distance of periastron $r_{\rm peri}$ produces similar but higher amplitude structures.
\item Due to the radial-drift of the dust content of the disc, for the very same flyby parameters the outcomes for the gas and for the dust are different. The dust disc is more compact and hence is less perturbed during the flyby than the gas. This leads to an increase of the dust-to-gas ratio accompanied by dust mixing within the disc. Moreover, the tidal interaction can potentially shift the inner pressure maxima in the disc outwards, especially for prograde encounters.
\item Due to gas drag, dust grains of different sizes (1 $\mu$m - 10 cm) exhibit different distributions for the same set of flyby parameters (cf. Figs.~\ref{fig:dustr45} and \ref{fig:dustr135}). Thus, the grains with Stokes numbers close to $1$ ($\sim$ 1 mm in this study) are most efficiently trapped along the spiral pressure maxima.
\item For inclined encounters ($\beta=45\degree, 90\degree, 135\degree$), we find that the disc is significantly warped --- both tilted and twisted. Warping increases with increasing $q$ and decreasing $r_{\rm peri}$, as expected. Flyby-induced warps can induce misalignments of tens of degrees. We find warping to be most efficient for inclined retrograde orbits (cf. Figs.~\ref{fig:tilttwist} and \ref{fig:tiltq}).
\item During the encounter, we find that the accretion rate of the primary star can be suddenly increased by an order of magnitude or more in time-scales of the order of $10$ yr. This translates into an increase in luminosity that may be linked to FU-Orionis events.
\end{enumerate}
A molecular cloud is a disturbing place to live.	 
 
\section*{Acknowledgements}
N.C. acknowledges financial support provided by FONDECYT grant 3170680. GD acknowledges financial support from the European Research Council (ERC) under the European Union's Horizon 2020 research and innovation programme (grant agreement No 681601). DJP and CP acknowledge funding from the Australian Research Council via FT170100040, FT130100034 and DP180104235. 
JC acknowledges support from CONICYT-Chile through FONDECYT (1141175) and Basal (PFB0609) grants. 
JC, PPP, and MM acknowledge financial support from ICM (Iniciativa Cient\'ifica Milenio, Chilean Ministry of Economy) via grant RC130007, the N\'ucleo Milenio de Formaci\'on Planetaria grant. MM acknowledge financial support from the CHINA-CONICYT fund, 4th call. FM\'e, GL and CP acknowledge funding from ANR of France under contract number ANR-16-CE31-0013. DJP and DM thank Pontificia Universidad Cat\'olica de Chile and the Millenium Discs Nucleus for hospitality and funding during a visit to Santiago where this project was initiated. This project was partially supported by the IDEXLyon project (contract n$\degr$ANR-16-IDEX-0005) under University of Lyon auspices. Figs.~\ref{fig:gas-panel}--\ref{fig:kinematics}, \ref{fig:onevstwo}, \ref{fig:r45q-dens} and \ref{fig:r45-kin} were made with {\sc splash} \citep{Price2007}. The Geryon2 cluster housed at the Centro de Astro-Ingenier\'ia UC was used for the calculations performed in this paper. The BASAL PFB-06 CATA, Anillo ACT-86, FONDEQUIP AIC-57, and QUIMAL 130008 provided funding for several improvements to the Geryon/Geryon2 cluster. This research used the ALICE High Performance Computing Facility and the DiRAC Data Intensive service operated by the University of Leicester IT Services. These resources form part of the STFC DiRAC HPC Facility (\url{www.dirac.ac.uk}), jointly funded by STFC and the Large Facilities Capital Fund of BIS via STFC capital grants ST/K000373/1 and ST/R002363/1 and STFC DiRAC Operations grant ST/R001014/1. DiRAC is part of the National e-Infrastructure. We thank Mohamed Hibat Allah for helping with the flyby setup at an early stage of this work. This work made extensive use of the NASA ADS database.

\bibliographystyle{mnras}
\bibliography{flybybiblio}

\appendix

\section{Initial setup for a flyby}
\label{sec:flybysetup}

We assume the flyby follows a parabolic orbit. We set the initial distance to be $10\,r_{\rm peri}$, where $r_{\rm peri}$ is the distance of closest approach (periastron). In what follows, when setting the initial positions and velocities, we assume that the orbit is in the \textit{xy}-plane. Then we rotate the initial position and initial velocity vectors to match the required inclination. Thus, the initial position ($x_i$,$y_i$,$z_i$) is given by
\begin{align}
   x_i &= -2 \sqrt{9} r_{\rm peri}, \\
   y_i &= r_{\rm peri} \left[1-{\left(\frac{x_i}{2r_{\rm peri}}\right)}^2\right], \\
   z_i &= 0.
\end{align}
To ensure a parabolic orbit, we set the initial velocity ($u_i$,$v_i$,$w_i$) to
\begin{align}
   u_i &= \left(1 + \frac{y_i}{10 r_{\rm peri}} \right)
   \sqrt{\frac{M_{\rm t}}{2r_{\rm peri}}}, \\
   v_i &= -(x_i/r_i)\sqrt{\frac{M_{\rm t}}{2r_{\rm peri}}}, \\
   w_i &= 0,
\label{eq:velflyby}
\end{align}
where $M_{\rm t}$ is the sum of the stellar masses.

We run each calculation such that the final distance between the stars is again $10 \, r_{\rm peri}$ (on the opposite side of the disc). To calculate this time we use Barker's equation:
\begin{equation}
   \Delta t = \sqrt{\frac{2 r_{\rm peri}^3}{G M_{\rm t}}} \left(D_f +
   \frac{1}{3} D_f^3 - D_i - \frac{1}{3} D_i^3 \right),
\end{equation}
where $D_{i,f} = \tan(\nu_{i,f}/2)$, and $\nu_{i,f} = \pi - \arctan(x_{i,f}/y_{i,f})$, is the true anomaly. Subscripts $i$ and $f$ refer to the initial and final position, respectively.

\section{One- vs two-fluid method comparison}
\label{sec:1vs2fluid}

\begin{figure}
\begin{center}
\includegraphics[width=0.5\textwidth,trim={2.7cm 1.45cm 0 0},clip]{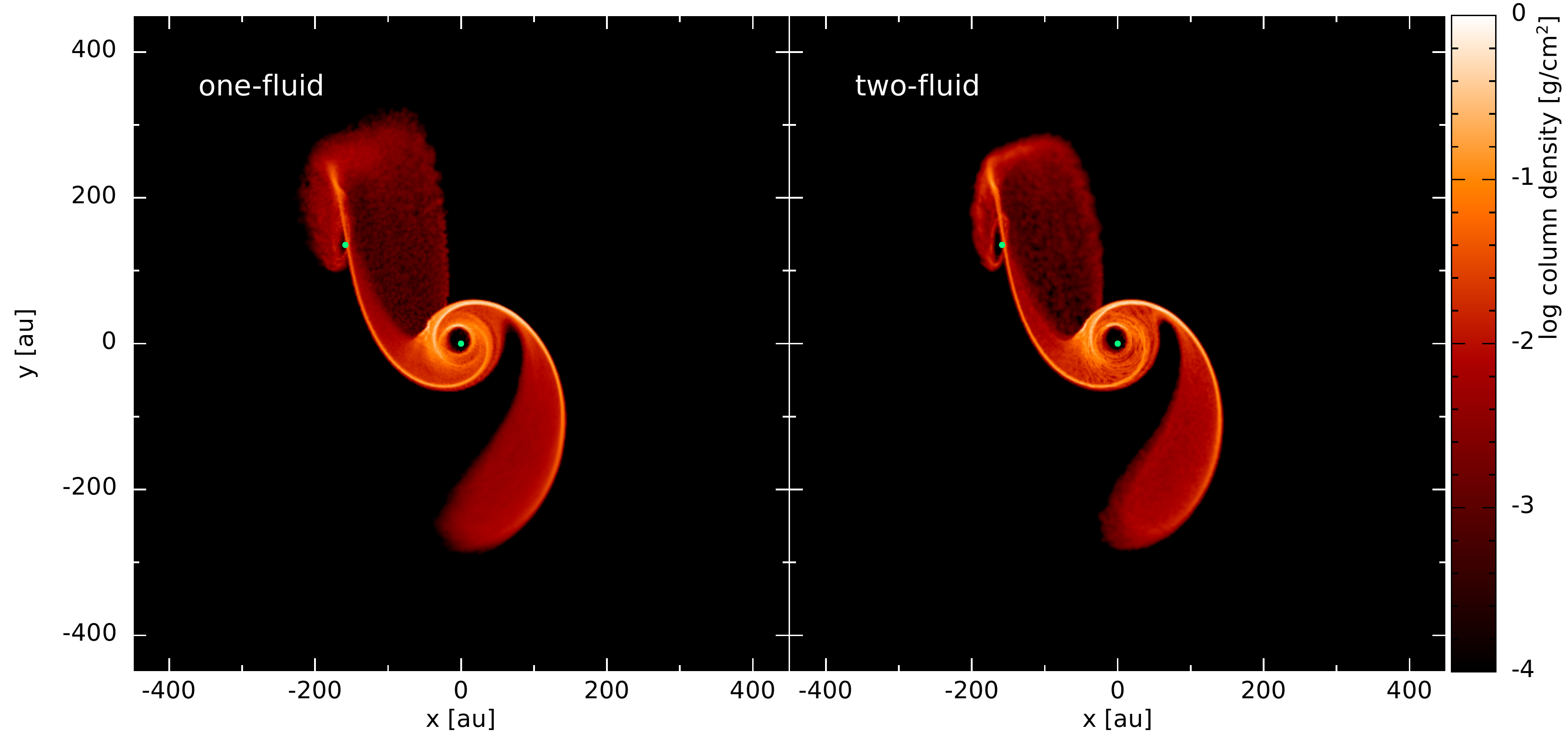}
\caption{Face-on view of the dust column density for $s=0.1$ mm and $\beta$45, using two different methods for computing the dust evolution. The snapshots correspond to the one-fluid (left) and the two-fluid (right) methods, and are taken shortly after the pericentre when the disc structure is the more complex. The disc rotation is anticlockwise. Both methods give very similar results, which validates our choice in Sect.~\ref{sec:dustmethod}.}
\label{fig:onevstwo}
\end{center}
\end{figure}

 Figure~\ref{fig:onevstwo}, presents a comparison of the dust distribution obtained with the one-fluid and the two-fluid methods, for $s~=~0.1$ mm and $\beta=45\degr$. Regardless of the method chosen to simulate the evolution of the dust density, we find visually identical disc structures. This indicates that the sharpness of the structures observed for different grain sizes is not due to the choice of the numerical method. Indeed, comparing the dust distribution for $s=10$ $\mu$m and $s=0.1$ mm using the one-fluid method, we observe that the spirals are narrower for the larger grains. Hence, this test proves that the more diffuse appearance of the $10\, \mu$m-spirals compared to the $0.1$ mm ones is physical. This sanity check justifies our choice of method for different grain sizes made in Section~\ref{sec:dustmethod}.

\section{Inner cavity and sink size}
\label{sec:cavitysink}

We test our assumption that the accretion radius of the sink size does not affect the structure of the inner regions of the disc. To do so, we consider two sizes for the sink particles ($1$ and $10$ au) and two resolutions ($10^6$ and $10^5$ SPH particles). In SPH, the resolution corresponds to the number of particles. For computational reasons, we performed this test only for one orbital inclination: $\beta180$, i.e. the case for which the inner cavity is the largest in our sample (cf. Fig.~\ref{fig:sigmavsr}). Recall that, given the large number of simulations carried out for this study, the choice of 10 au for the accretion radius was purely for numerical convenience.

\begin{figure}
\begin{center}
\includegraphics[width=0.48\textwidth]{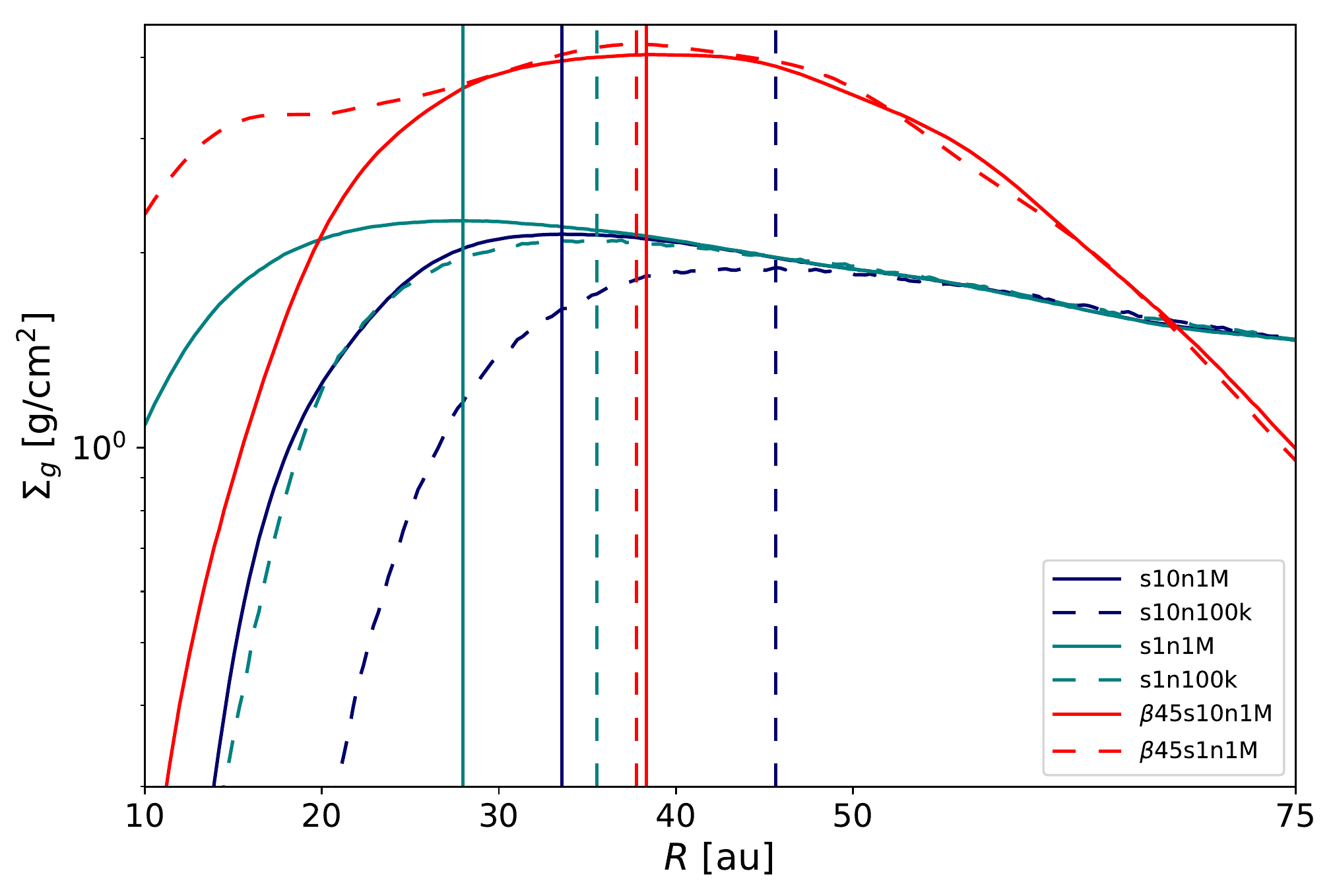}
\caption{Azimuthally averaged gas surface density 2\,700 yr after the flyby for $\beta180$. The curves correspond to different resolutions and accretion radii for the sink. In the legend, the prefix ``s'' stands for the accretion radius in au; while ``n'' stands for the number of gas particles considered. For completeness we also show the gas surface density after 2\,700 yr for $\beta$45 for $10^6$ and $10^5$ particles. The lower the resolution and/or the larger the accretion radius, the larger the inner cavity (as expected). The vertical lines correspond to the location of the surface density maxima (from the inside out): 28.0, 33.6, 35.5, 37.8, 38.3 and 45.6 au, using the same convention as in the legend. For the parameters considered in this study, our results are marginally affected by these effects.}
\label{fig:sinktest}
\end{center}
\end{figure}

In Fig.~\ref{fig:sinktest}, the prefixes s1 and s10 correspond to sink sizes equal to $1$ and $10$ au (respectively); while n1M and n100k stand for $n=10^6$ and $10^5$ (respectively), where $n$ is the number of gas particles. By comparing s1n1M and s10n1M, we see that a larger accretion radius for the sink particle empties the inner regions of the disc more efficiently. This effect is purely numerical and only affects the surface density below $30$ au. However, despite the dramatic drop in density observed for s10n1M, the radial location of the density maximum is practically the same for s1n1M and s10n1M: 28.0 and 33.6 au, respectively. Reducing the sink size by a factor 10 only shifts the location of the density maximum by 5.6 au.

Additionally, lowering the resolution from $10^6$ to $10^5$ changes the surface density of the inner regions: the lower the resolution, the lower the density. The difference in location between s1n1M and s1n100k is equal to $7.5$ au. Lower resolutions artificially shift the radial location towards the outer regions of the disc ($>30$ au for s1n100k). This unavoidably affects the dust distribution of the inner regions of the disc. Remarkably, increasing the accretion radius of the sink by a factor of $10$ is equivalent to lowering the resolution by the same factor (s10n1M and s1n100k). Beyond $40$ au, all the surface density profiles overlay on top of each other, regardless the resolution and/or the accretion radius. For completeness, we also show the gas surface density for $\beta45$ considering two different value for the sink radius ($1$ and $10$ au). In this particular case, the radial location of the peak value is sensibly the same ($37.8$ and $38.3$ au).

Hence, our choice of an accretion radius of 10 au for the sink particle does not dramatically affect the dust distribution beyond 30 au. For the resolution considered in this study ($10^6$), higher accretion radii for the sink ($>10$ au) would produce artificially large inner cavities. Finally, a lower accretion radius --- more realistic, but also more expensive --- would produce smoother profiles with higher gas densities in the inner regions. However, in our flyby simulations the most prominent features appear in the outer regions of the disc where the accretion radius of the sink has no effect.

\begin{figure}
\begin{center}
\includegraphics[width=0.48\textwidth]{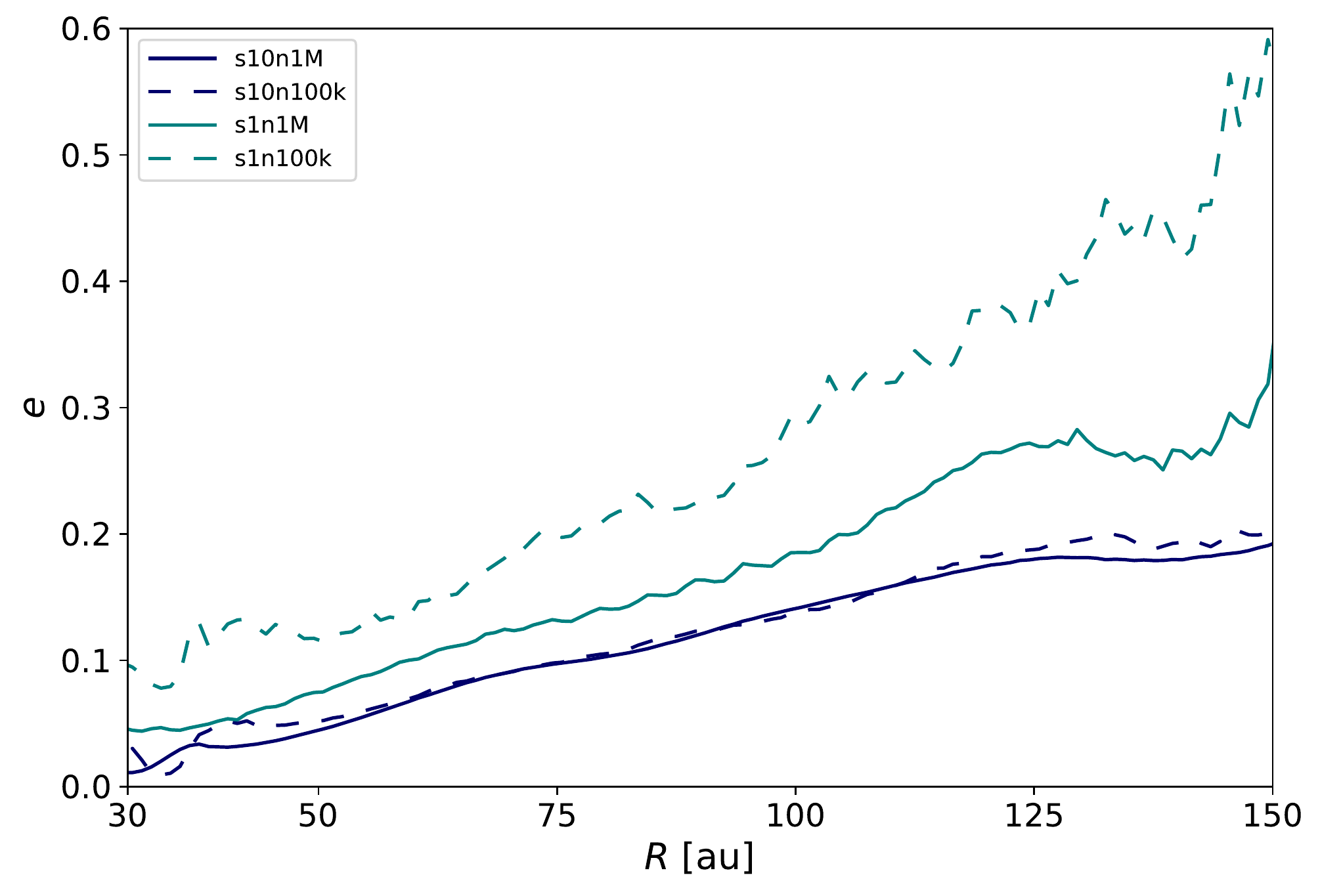}
\caption{Eccentricity as a function of the radial distance to the star, 2\,700 yr after the flyby for $\beta180$. As in Fig.~\ref{fig:sinktest}, we show different sink sizes and resolutions.}
\label{fig:sinktest-ecc}
\end{center}
\end{figure}

In Figure~\ref{fig:sinktest-ecc} we show the eccentricity as a function of the distance to the star for the same runs shown in Fig.~\ref{fig:sinktest}. The size of the sink particle has a minor effect on the eccentricity compared to resolution effects. For a fixed resolution, a larger sink size translates into a more efficient eccentricity damping in the disc. Additionally, for a fixed sink size, a lower resolution leads to higher eccentricities throughout the disc (specially in the inner regions). For the denser regions (i.e. beyond the maximum density location at roughly 30 au) and for high resolution, the eccentricity is sensibly the same. Therefore, the results presented in this study are marginally dependent on the resolution and the sink size for the outer disc regions.

\begin{figure}
\begin{center}
\includegraphics[width=0.48\textwidth]{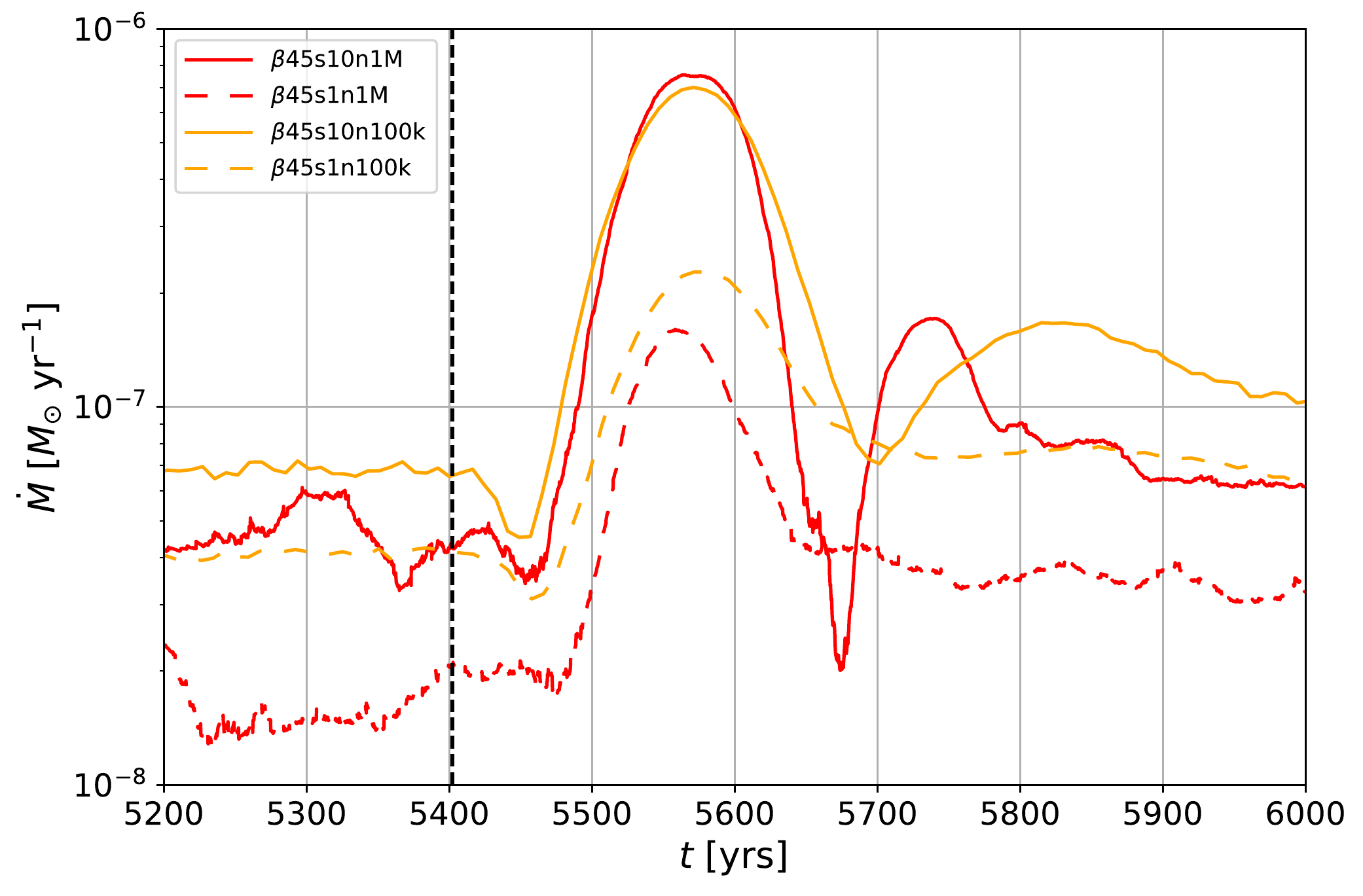}
\caption{Accretion rates onto the primary star during the encounter for $\beta45$ for different sink radii: 10 and 1 au in solid and dashed lines respectively. The vertical dashed line corresponds to the passage at periastron. Decreasing the sink radius by a factor of $10$ reduces the accretion rate by a factor of $3$ roughly.}
\label{fig:sinktest-Mdot}
\end{center}
\end{figure}

Finally, in Fig.~\ref{fig:sinktest-Mdot} we show that reducing the sink radius by a factor of $10$ is equivalent to dividing the accretion rate by $3$ approximately. We perform this test for the case exhibiting the strongest accretion event, i.e. $\beta45$. The accretion rates obtained for lower resolution ($10^5$) exhibit a similar trend. As expected, a larger sink radius {and/or a lower resolution} lead to higher accretion rates. However, the timespan of the event and the shape of the curves are comparable in all the runs. Remarkably, the ratios between the peak values of accretion and the values before the passage at pericentre are comparable. Therefore, the accretion event itself is not a numerical artefact and constitutes a distinctive signature of a flyby. Regardless of the sink size and the resolution, we observe a flyby-induced accretion event for prograde configurations. Although the accretion rate value itself depends on the numerical setup, accretion events are an expected dynamical outcome of prograde encounters. We recall that retrograde encounters as $\beta180$ do not show such accretion events.

\section{Mass ratio effects}
\label{sec:appendix-kin-q}

\begin{figure*}
\begin{center}
\includegraphics[width=0.93\textwidth]{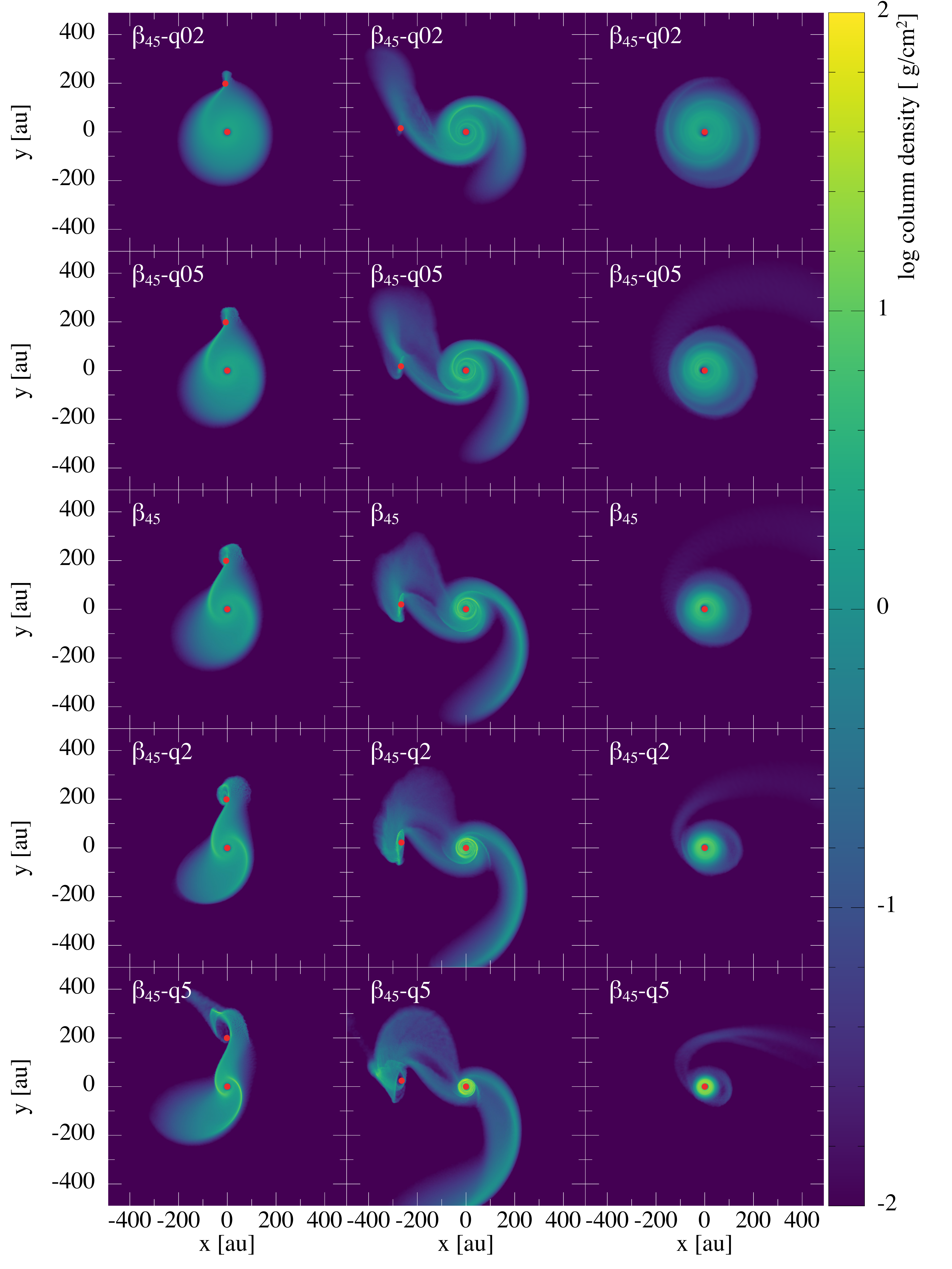}
\caption{Face-on view of the gas column density for (from top to bottom rows): $\beta$45q02, $\beta$45q05, $\beta$45, $\beta$45q2 and $\beta$45q5. From left to right columns: $t=5\,400$ yr (periastron), $t=5\,950$ yr and $t=8\,100$ yr. The disc rotation is anticlockwise. Sink particles (in red) are large for visualization purposes only. The higher the mass ratio $q$, the smaller the disc final size. The morphology of the spirals is similar, but sharper and more extended for high values of $q$.}
\label{fig:r45q-dens}
\end{center}
\end{figure*}

\begin{figure*}
\begin{center}
\includegraphics[width=0.93\textwidth]{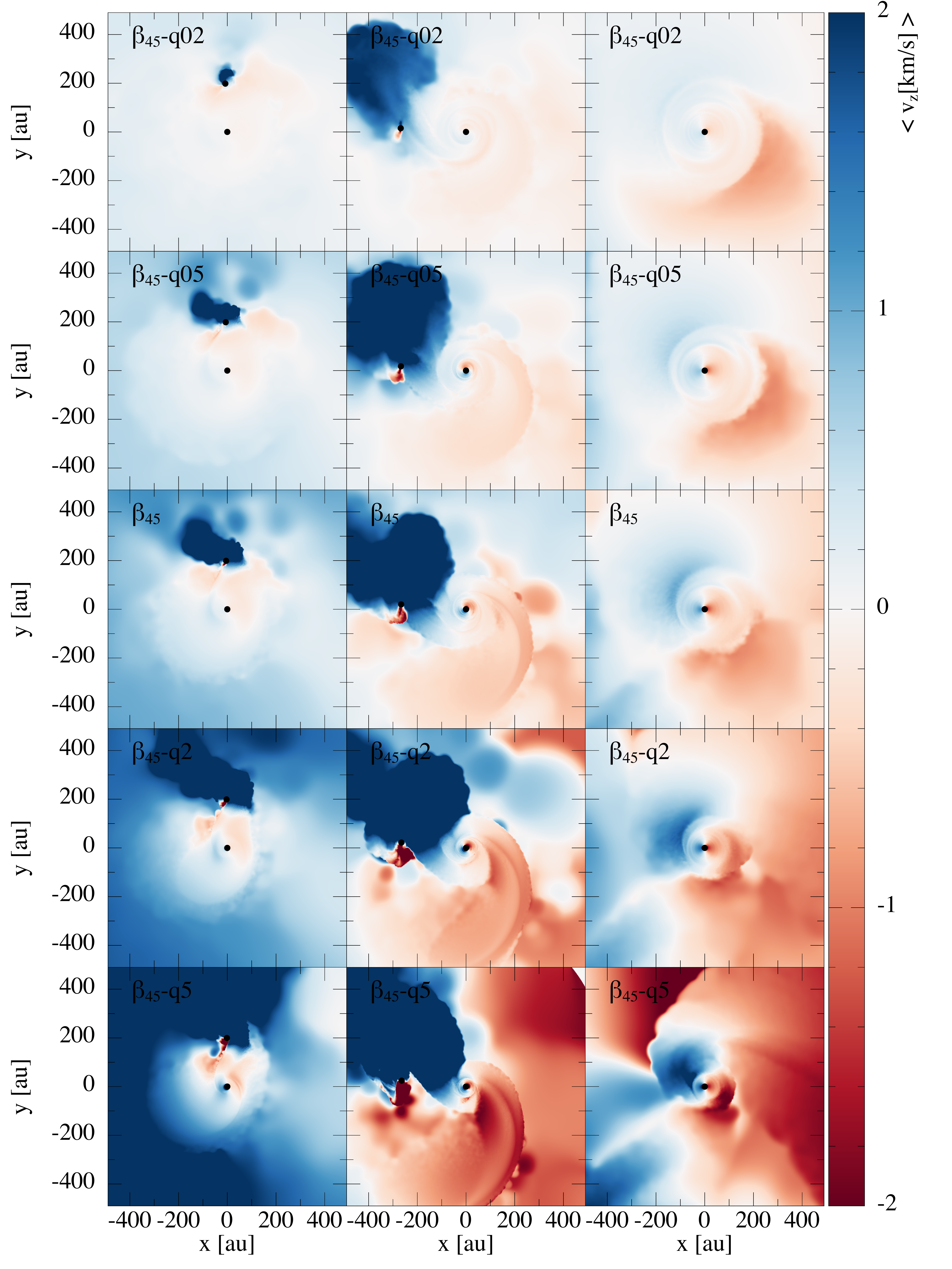}
\caption{Vertical velocity integrated along the line of sight. From top to bottom: $\beta$45q02, $\beta$45q05, $\beta$45, $\beta$45q2 and $\beta$45q5. From left to right columns: $t=5\,400$ yr (periastron), $t=5\,950$ yr and $t=8\,100$ yr. The disc rotation is anticlockwise. The higher the mass ratio $q$, the higher the vertical velocity departures. Even low-mass companions (as $q=0.2$) produce significant kinematic signatures.}
\label{fig:r45-kin}
\end{center}
\end{figure*}

\begin{figure}
\begin{center}
\includegraphics[width=0.5\textwidth]{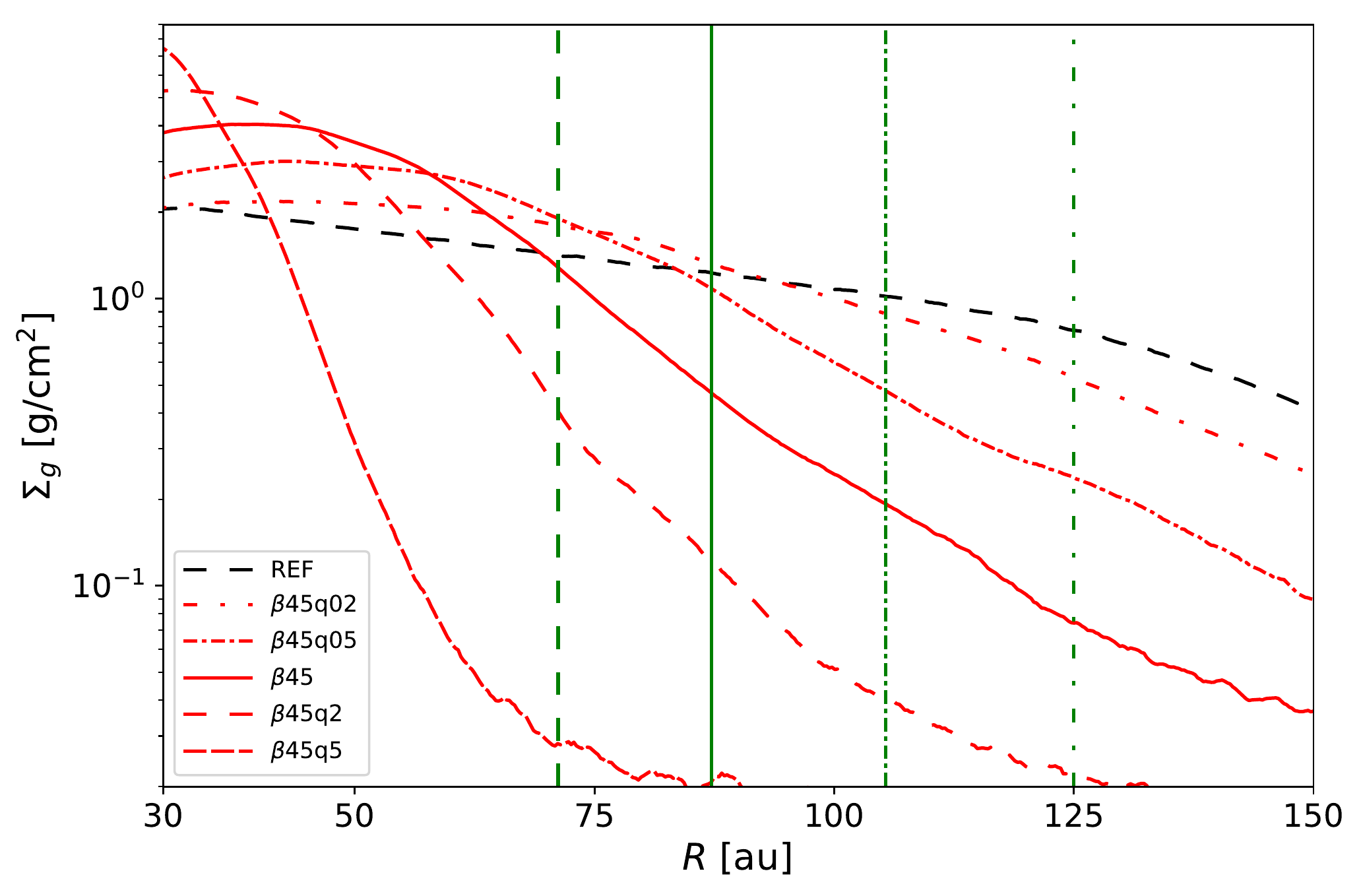}
\caption{Azimuthally averaged gas surface density 2\,700 yr after the flyby for $\beta=45\degree$ and different values of $q$ (red): 0.2 (dotdashed), 0.5 (densely dotdashed), 1 (solid), 2 (dashed) and 5 (densely dashed). The NoFB case is represented by the (dashed) black line. The vertical green lines correspond to the disc sizes obtained with the ``63\% disc mass'' definition (cf. Table~\ref{tab:rd}).}
\label{fig:gasdensq}
\end{center}
\end{figure}

In Fig.~\ref{fig:r45q-dens} we show three snapshots of simulations with $\beta=45\degree$, but with different values of $q$. We see that the morphology of the spirals is similar for all the cases considered. The spiral on the right becomes sharper (i.e. narrower) and more radially extended for increasing $q$. This is because the gravitational perturbation is stronger and the encounter occurs at higher velocity (see Eqs.~\ref{eq:velflyby}). A similar trend is observed in the kinematic field of the disc shown in Fig.~\ref{fig:r45-kin}. The higher the mass ratio, the strongest the vertical deviation in the velocity field of the disc. While these barely reach values of $1$ km/s for $q=0.2$, for $q=5$ the blue- and red-shifted regions have several km/s towards and away from the observer (respectively). Last, the disc is more heavily truncated for high values of $q$ as shown Fig.~\ref{fig:gasdensq}. For completeness, we plot the disc radii reported in Table~\ref{tab:rd} using the same definition as \cite{Bate2018}.

\begin{figure}
\begin{center}
\includegraphics[width=0.5\textwidth]{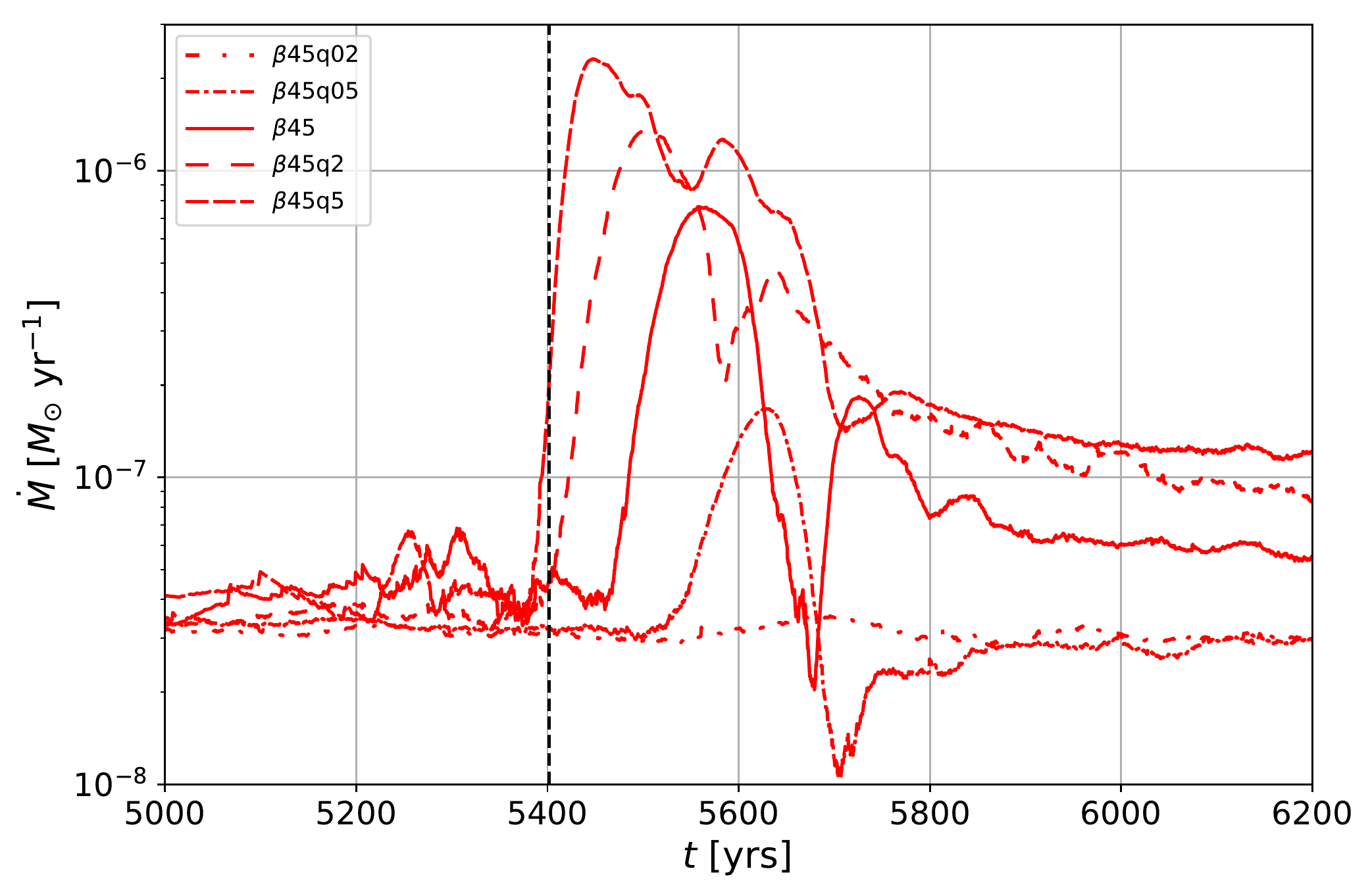}
\caption{Accretion rates for flyby simulations with $\beta=45\degree$ and different values of $q$: 0.2 (dotdashed), 0.5 (densely dotdashed), 1 (solid), 2 (dashed) and 5 (densely dashed). Higher values of $q$ produce more significant and faster accretion events.}
\label{fig:accretion-q}
\end{center}
\end{figure}

Additionally, increasing the value of $q$ also increases the amount of matter accreted by the central star. This translates into an increase in accretion rate as shown in Fig.~\ref{fig:accretion-q}. Also, the increase in luminosity after the periastron occurs faster for high values of $q$: $\sim 200$ yr for $\beta$45q05 and $\sim 25$ yr for $\beta$45q5. The higher $q$, the brighter the star and the faster the outburst.

\begin{figure}
\begin{center}
\includegraphics[width=0.5\textwidth]{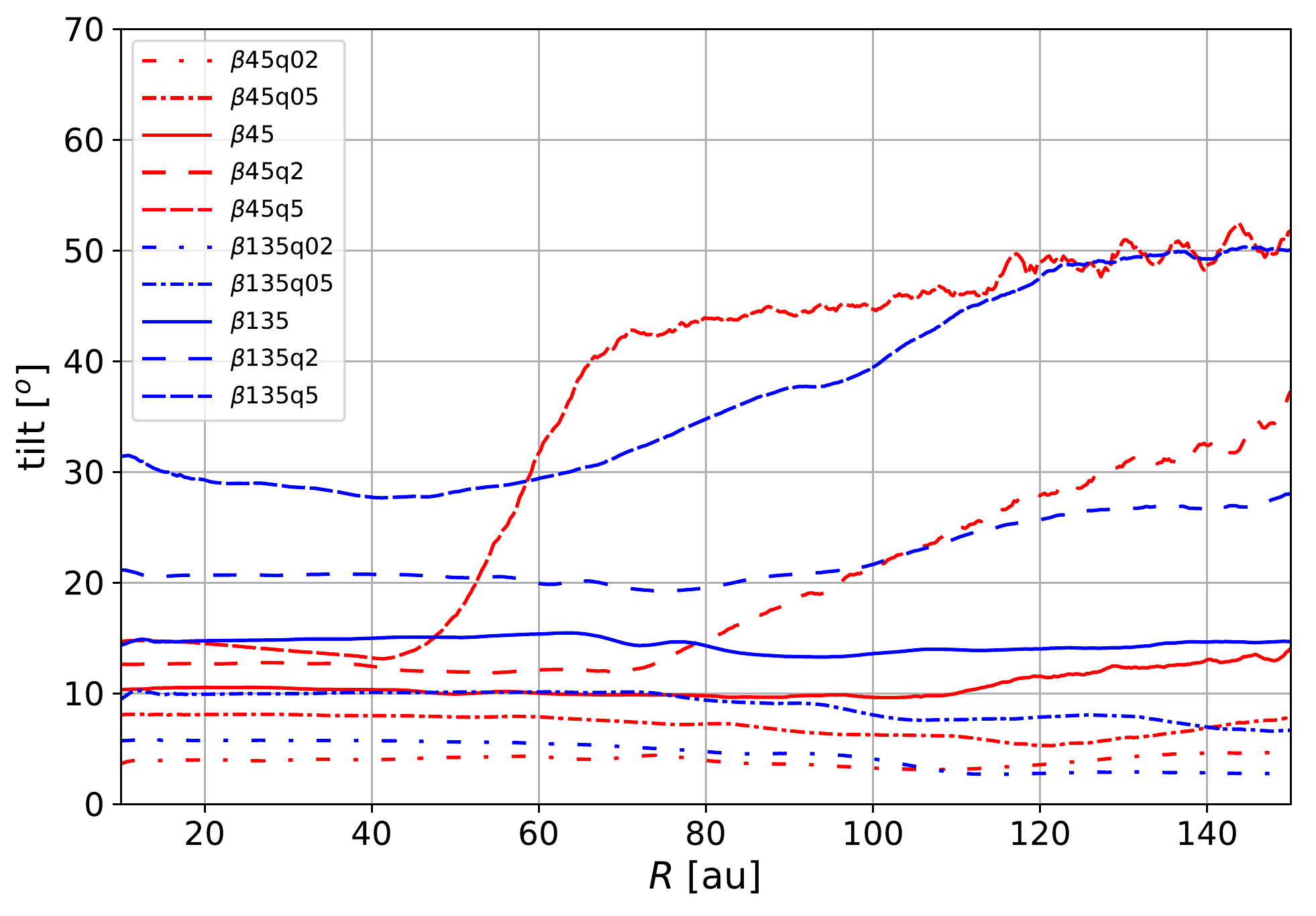}
\caption{Tilt of the disc 2\,700 yr after the flyby for $\beta$45 (red) and $\beta$135 (blue) for different mass ratios: 0.2 (dashdotted), 0.5 (densely dashdotted), 1 (solid), 2 (dashed), 5 (densely dashed). The tilt increases with $q$ for a fixed perturber's orbital inclination. Inclined retrograde orbits efficiently tilt the disc by several tens of degrees.}
\label{fig:tiltq}
\end{center}
\end{figure}

\begin{figure}
\begin{center}
\includegraphics[width=0.5\textwidth]{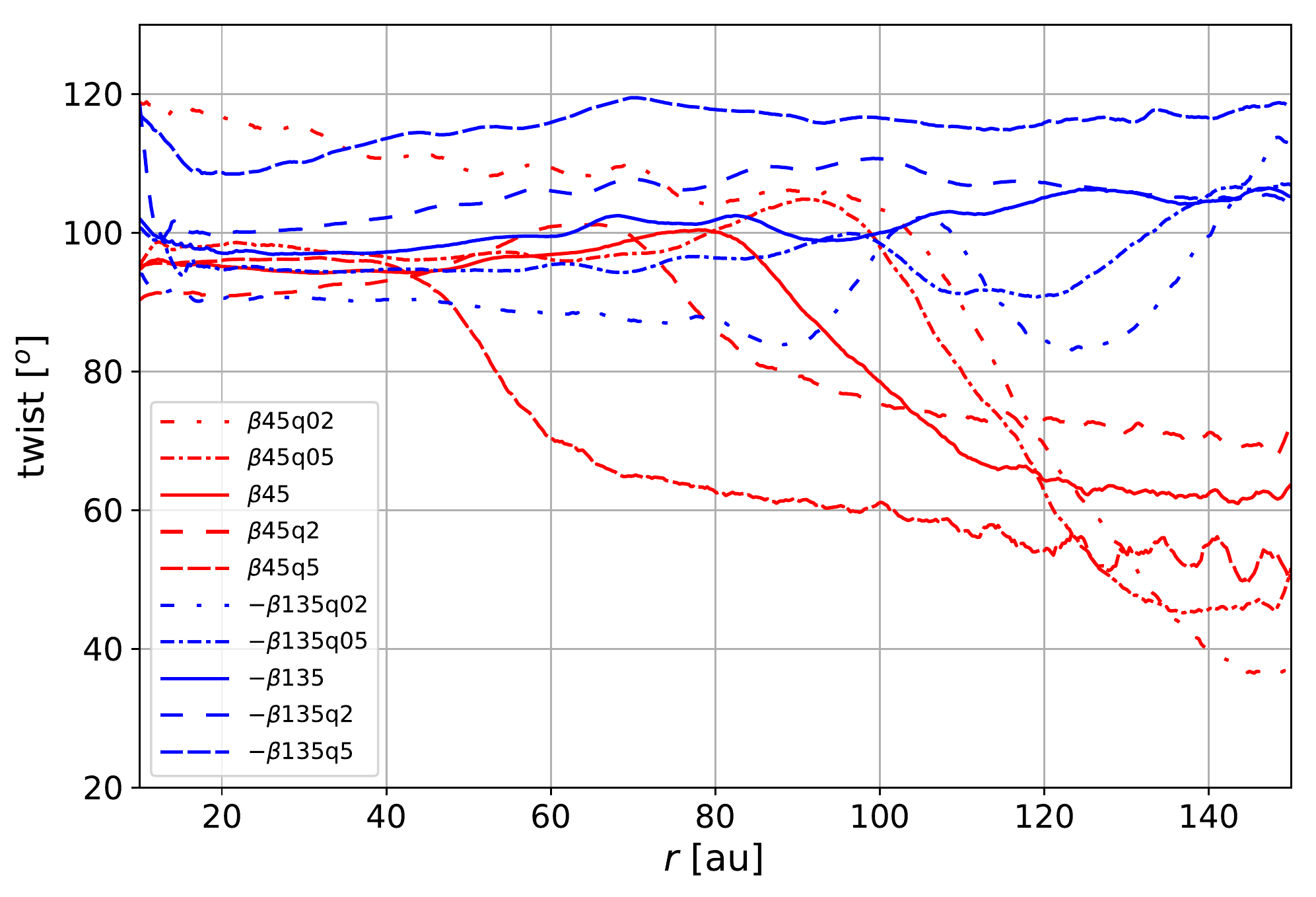}
\caption{Twist of the disc 2\,700 yr after the flyby for $\beta$45 (red) and $\beta$135 (blue) for different mass ratios: 0.2 (dashdotted), 0.5 (densely dashdotted), 1 (solid), 2 (dashed), 5 (densely dashed). Despite the broad range in $q$, the values of the twist for the bulk of the disc (below 50 au) are comprised between $90\degree$ and $110\degree$.}
\label{fig:twistq}
\end{center}
\end{figure}

Finally, in Fig.~\ref{fig:tiltq} and Fig.~\ref{fig:twistq} we show the tilt and the twist of the disc (respectively), 2\,700 yrs after the flyby. We see that the same trends reported in Sect.~\ref{sec:warps} hold. As expected, the tilt increases with $q$. Inclined retrograde orbits efficiently tilt the disc by several tens of degrees as $\beta$135q5. Remarkably, even by changing the mass ratio by a factor of 25 (from 0.2 to 5), the twist of the disc remains comprised between $90\degree$ and $110\degree$, for both orbital inclinations.

\label{lastpage}
\end{document}